\documentclass[english,aps,prb,10pt,twocolumn,superscriptaddress,showpacs]{revtex4-1}
\usepackage[T1]{fontenc}
\usepackage[latin9]{inputenc}

\usepackage{graphicx,wrapfig}
\usepackage{babel}
\usepackage{hyperref}

\usepackage{color}

\makeatletter
\@ifundefined{textcolor}{}
{%
 \definecolor{BLACK}{gray}{0}
 \definecolor{WHITE}{gray}{1}
 \definecolor{RED}{rgb}{1,0,0}
 \definecolor{GREEN}{rgb}{0,1,0}
 \definecolor{BLUE}{rgb}{0,0,1}
 \definecolor{CYAN}{cmyk}{1,0,0,0}
 \definecolor{MAGENTA}{cmyk}{0,1,0,0}
 \definecolor{YELLOW}{cmyk}{0,0,1,0}
 }

\newcommand{\bea}{\begin{eqnarray}}
\newcommand{\eea}{\end{eqnarray}}
\newcommand{\la}{\label}
\newcommand{\be}{\begin{equation}}
\newcommand{\ee}{\end{equation}}

\makeatother

\begin{document}

\title{Hydrodynamics of cold atomic gases in the limit of weak nonlinearity, \\ dispersion and dissipation
}

\author{Manas Kulkarni}
\affiliation{Department of Physics, University of Toronto, Toronto, Ontario M5S 1A7, Canada}
\affiliation{Chemical Physics Theory Group, Department of Chemistry,
University of Toronto, Toronto, Ontario M5S 3H6, Canada}

\author{Alexander G. Abanov}
\affiliation{Department of Physics and Astronomy, Stony Brook University,
Stony Brook, NY 11794-3800, USA}

\date{\today}

\begin{abstract}

Dynamics of interacting cold atomic gases have recently become a focus of both experimental and theoretical studies. Often cold atom systems show hydrodynamic behavior and support the propagation of nonlinear dispersive waves. Although this propagation depends on many details of the system, a great insight can be obtained in the rather universal limit of weak nonlinearity, dispersion and dissipation (WNDD). In this limit, using a reductive perturbation method we map some of the hydrodynamic models relevant to cold atoms to well known chiral one-dimensional equations such as KdV, Burgers, KdV-Burgers, and Benjamin-Ono equations.  These equations have been thoroughly studied in literature. The mapping gives us a simple way to make estimates for original hydrodynamic equations and to study  the interplay between nonlinearity, dissipation and dispersion which are the hallmarks of nonlinear hydrodynamics.


\end{abstract}

\maketitle

\setcounter{tocdepth}{1}
\tableofcontents

\section{Introduction}
\label{intro}

Recently, there has been a flurry of experiments that study the collective behavior of systems of particles with both bosonic \cite{dutton,Khaykovich,Simula,meppilink} and fermionic\cite{hara-science,key-4thomas,key-5thomas1,key-6thomas2,d-jin,jjka,jjka1,randy2012} statistics. There
have also appeared experimental realizations of cold atom systems with long range interatomic interactions (such as dipolar bosons \cite{dy,axel,axelexp}, dipolar fermions\cite{debi}, ions, and Rydberg atoms). Investigation of dynamics of such systems is interesting and important for many reasons. The dynamics can provide an important information about the nature and strength of particle interactions. In many cases the nonlinear dynamical evolution leads to a formation of shock waves and solitons. Out of equilibrium dynamics in the presence of defects and disorder (see, e.g., Ref. \onlinecite{randy}) can shine light on dissipation and localization phenomena. One of the most intriguing and important direction is the search for universality in dynamical properties of systems governed by different microscopic Hamiltonians.\cite{eugene2012}

The dynamics of cold Bose gases is being extensively investigated both theoretically\cite{BD,AblowitzEtAl} and experimentally. Experiments on Bose-Einstein condensates (BEC) include sound velocity measurements \cite{Simula} and collisions of atomic clouds which lead to the observation of dispersive shock waves and soliton trains\cite{EngelsFirstObservation,meppilink}. A dissipative transport in BEC has been recently investigated in Ref.~\onlinecite{randy}. The propagation of matter-wave soliton trains in BECs has been realized in Ref.~\onlinecite{randy1}. A strongly interacting limit of interacting 1d bosons a.k.a. the Tonk's gas analogous to free fermions has been realized in Ref.~\onlinecite{wiess_tonks}.

The dynamics of Fermi systems\cite{wolfgang} has also been of great interest recently\cite{mz,JosephSound,jjka}. For instance, in Ref.~\onlinecite{mz} the spin transport was studied by colliding two oppositely spin-polarized clouds of fermions. In Refs.~\onlinecite{jjka,jjka1} some aspects of nonlinear hydrodynamics (such as shock waves) in unitary Fermi gases were investigated. These experiments have also motivated new numerical studies of nonlinear dynamics in Fermi systems\cite{aurel}.

It is also worth mentioning that in addition to conventional Fermi and Bose many body systems there exists a remarkable family of models interpolating between fermionic and bosonic systems. This is the family of Calogero models. The exact nonlinear collective description for these models is known\cite{JevickiSakita,Jevicki-1992,2009-AbanovBettelheimWiegmann,2005-AbanovWiegmann} and has a form of equations of hydrodynamic type. These hydrodynamic equations are integrable and exhibit features such as solitons and dispersive shock waves. The equations are related to a known integro-differential equation -- the Benjamin-Ono equation\cite{Jevicki-1992,2009-AbanovBettelheimWiegmann,2005-AbanovWiegmann}. Unlike other integrable models, the Calogero model and its hydrodynamic description retain integrability even in the presence of an external harmonic potential. In particular the solutions of multi-soliton type have been found for these models in Ref.~\onlinecite{harmonic_soliton}.

Because of the variety of systems and models used in studying nonlinear dynamics of cold atom systems the understanding of major effects resulting from the interplay between nonlinearity, dispersion and dissipation within a simple unifying description would be very useful. Fortunately, this description is well known in the limit of weak nonlinearity, dispersion, and dissipation. It is summarized by the Korteweg-de Vries-Burgers equation (KdVB) (see eq.~\ref{kdvb} below). The main goal of this paper is to connect this universal picture to some particular models used in collective descriptions of cold atom systems. For simplicity we concentrate here on simple Galilean invariant fluids although many features of nonlinear dynamics described in this paper can be found in more complex systems as well.

\begin{figure}
\includegraphics[scale=.35,angle=0]{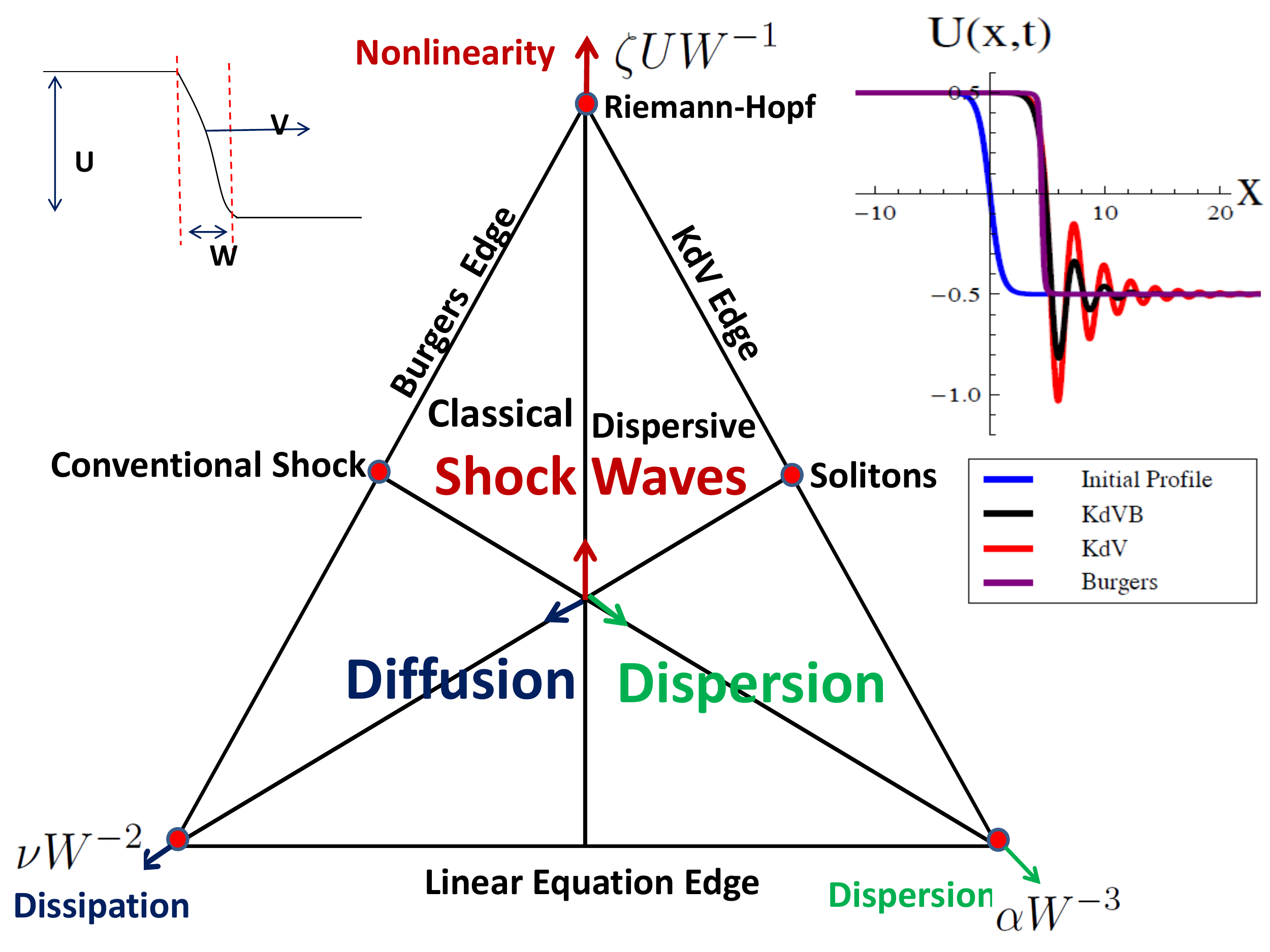}
\caption{A schematic phase-diagram summarizing various hydrodynamic regimes of KdVB equation. We consider a typical step profile shown in the left insert characterized by a typical width $W$ and an amplitude $U$.
The directions towards the corners  of the triangle correspond to the increase of corresponding $\nu$, $\alpha$ and $\zeta$-terms of the KdVB equation (\ref{kdvb}) evaluated on the typical profile. Corresponding to the strength of those terms the diagram is divided into regions corresponding to diffusive, dispersive and shock wave regimes with dominating dissipation, dispersion and nonlinearity terms, respectively. The edges of the triangle correspond to the exactly solvable equations with one of the terms vanishing. See sections \protect\ref{sec:redpert},\protect\ref{sec:dsw} for details. }
\label{estimate}
\end{figure}

The following is a brief outline of this paper. We start by constructing a rather general one-dimensional model of a simple Galilean invariant fluid in Sec.~\ref{sec:hm}. We linearize this model in Sec.~ \ref{sec:lin} and proceed to the derivation of the effective KdVB using the reductive perturbation method in Sec.~\ref{kdvb_derivation}. We describe the effects of nonlinearity, dispersion and dissipation for the effective KdVB equation and introduce corresponding scales and limits. These scales and limits are presented in the triangle phase diagram in Fig.~\ref{estimate}. We conclude the qualitative description of KdVB dynamics in Sec.~\ref{sec:dsw} describing two different scenaria for shock wave formation in the presence of both dissipation and dispersion. Finally, in Sec.~\ref{examples} we give relations between the coefficients of the effective KdVB equation and parameters of some models used in cold atom studies. We conclude with some open questions and possible generalizations of our results in Sec.~\ref{conclusion} and describe non-KdVB universal behaviors for some systems with long range interactions in Appendix \ref{long-range-app}.

\section{Hydrodynamic model}
\label{sec:hm}

The goal of this section is to construct a hydrodynamic model (hydrodynamic equations) describing a rather general one-dimensional motion of a simple fluid. An effective one-dimensional motion usually describes a class of solutions of three-dimensional hydrodynamics obtained by a requirement on hydrodynamic fields $\rho(x,y,z;t)=\rho(x,t)$ etc. or through more complicated procedures of a dimensional reduction, e.g., due to the presence of a quasi-one-dimensional trap in cold atom systems (see e.g., Sec.~\ref{examples} and Refs. \onlinecite{jjka,jjka1}). An effective ``truly'' 1d model can also appear as a result of the transverse quantization in a quantum system\cite{BD,olshani1}. In the following we assume that the reduction to one dimension is already performed and outline other assumptions that we use to construct an effective one-dimensional hydrodynamic model.

\paragraph{A fluid is one-component.} We do not consider here more complicated theories of mixtures of fluids.  In particular, we do not consider the two-fluid hydrodynamics of superfluids with non-vanishing normal component. If a system under consideration is in the superfluid regime we assume here that the normal component can be ``integrated out'' resulting in a hydrodynamics of an effective single-component fluid with some additional terms (such as viscous terms) generated by the normal component.

\paragraph{Entropy generation is very small.} This assumption allows us to consider closed equations for density and velocity fields only and assume that the fluid motion is isentropic. Combined with the first assumption this means that the only relevant degrees of freedom can be described by one-dimensional velocity $v(x,t)$ and density $\rho(x,t)$ fields.

\paragraph{Locality.} We assume that the energy functional is local in density and velocity fields and equations of motion contain only local values of fields and their derivatives. This constraint can be relaxed (see examples in Appendix \ref{long-range-app}).
	
\paragraph{Galilean invariance.} In the presence of Galilean invariance the energy density of the fluid is given by $\rho v^{2}/2$ with other terms either independent of velocity or proportional to velocity gradients. Systems without this invariance cannot be described by a simple hydrodynamic Hamiltonian (\ref{hydroham}). We focus on Galilean invariant systems in this work for simplicity but do expect that the chiral differential equations of the form (\ref{kdvb}) still provide a rather universal effective description of collective dynamics even in the absence of Galilean invariance (e.g., for the effective description of XXZ spin chain in external magnetic field \cite{aditi_xxz}).
	
\paragraph{Energy functional contains density derivatives up to the second order.} This approximation is known as the Boussinesque approximation. In the absence of dissipation this assumption combined with previous assumptions leads to the following hydrodynamic Hamiltonian
\be
	H = \int dx\, \left[\frac{\rho v^{2}}{2} +\rho\epsilon(\rho)
	+A(\rho) \frac{(\partial_{x}\rho)^{2}}{4\rho}\right].
 \la{hydroham}
\ee
Here $A(\rho)$ is some function of density. We notice here that if $A=const$ the gradient term becomes the Madelung term well-known in superfluid hydrodynamics. The hydrodynamic Hamiltonian (\ref{hydroham}) equipped with Poisson's brackets
\begin{equation}
	\{\rho(x),v(x')\}=\partial_{x}\delta(x-x')
 \label{PBrhov}
\end{equation}
generates hydrodynamic equations of an ideal fluid \cite{landau1941,LandauFluids,variational}.

\paragraph{Dissipative function is quadratic in velocity gradients.} The positivity of the dissipative function does not allow for terms linear in velocity gradients and we have generically for (Rayleigh) dissipative function
\be
	F=\frac{1}{2}\int dx\, \eta_{B}(\rho)(\partial_{x}v)^{2}.
 \la{dissfun}
\ee
where $\eta_{B}(\rho)$ is a bulk viscosity of a 1D fluid.
In the presence of dissipation the hydrodynamic system is not Hamiltonian and Hamilton equations following from (\ref{hydroham},\ref{PBrhov}) should be supplemented by dissipative terms encoded in (\ref{dissfun}). We arrive at the following equations of motion
\bea
	\partial_{t}\rho &+& \partial_{x}(\rho v) = 0,
 \la{cont} \\
 	\partial_{t}v &+& \partial_{x}\left(\frac{v^{2}}{2}+w
	-(\partial_{\rho}A)(\partial_{x}\sqrt{\rho})^{2} -A\frac{\partial_{x}^{2}\sqrt{\rho}}{\sqrt{\rho}}\right)
 \nonumber \\
	&=& \frac{1}{\rho}\partial_{x}\left(\eta_{B}\partial_{x}v\right),
 \la{euler}
\eea
where $w=\partial_{\rho}(\rho\epsilon(\rho))$ is the specific enthalpy (same as the chemical potential at zero temperature) of the fluid and the r.h.s. of the second equation is obtained as $-\frac{1}{\rho}\frac{\delta F}{\delta v}$.
The rate of the dissipation of energy is then given by $2F$.

Let us rewrite these equation in terms of density and current (momentum density) $j=\rho v$. We have
\bea
	\partial_{t}\rho &+& \partial_{x}j = 0,
 \\
 	\partial_{t}j &+& \partial_{x}(T+T') = 0,
 \\
 	j &=& \rho v,
 \\
 	T &=& \rho v^{2} + P
	-\frac{1}{2}\rho\sqrt{A}\partial_{x}\Big(\sqrt{A}\, \partial_{x} \ln \rho  \Big),
 \la{T}
 \\
 	T' &=& -\eta_{B} \partial_{x}v.
 \la{Tprime}
\eea
Here the pressure $P(\rho)$ is given by $\partial_{\rho}P=\rho\partial_{\rho}w$. The last term of (\ref{T}) is called the ``quantum pressure term'' in the context of quantum fluids. The viscous part $T'$ of the stress tensor is linear in the gradient of velocity and is obtained from $\partial_{x}T'=\delta F/\delta v$\footnote{Equations with $T'$ are not Hamiltonian and correspond to the energy dissipation rate $2F$.}

Many models of recent interest can be casted in the form of the equations  (\ref{cont},\ref{euler}). The expressions for $w(\rho)$, $A(\rho)$ and $\eta_{B}(\rho)$ for some of those models are listed in the Table \ref{table_values}.

Equations (\ref{cont},\ref{euler}) are nonlinear. The nonlinearity results in the steepening of density and velocity profiles during the evolution and in shock waves. The quantum pressure terms (terms containing $A(\rho)$) depend on density gradients. They typically give rise to dispersive oscillations of densities and velocities. The viscosity term (the r.h.s. of (\ref{euler}) contains the gradient of velocity field. It results in damping. All these terms play rather different roles in fluid dynamics and understanding their interplay is of great interest both theoretically and experimentally.

\section{Linearization}
\label{sec:lin}

Let us start by studying the system (\ref{cont},\ref{euler}) in the limit of small deviations from the uniform solution $\rho(x)=\rho_{0}$, $v(x)=0$. We consider this solution as a background configuration of fields and linearize equations in $\delta \rho(x)=\rho(x)-\rho_{0}$ and $\delta v(x)=v(x)$. We obtain
\bea
	\delta\dot{\rho} &+& \rho_{0}\delta v'= 0,
 \la{cont-lin}\\
 	\delta \dot v &+& \frac{(\partial_{\rho}P)_{0}}{\rho_{0}}\delta\rho'
	-\frac{A_{0}}{2\rho_{0}}\delta\rho'''
 	= \frac{\eta_{B0}}{\rho_{0}}\delta v''.
 \la{euler-lin}
\eea
Here subscript ``0'' means that the quantity is calculated at $\rho=\rho_{0}$.
Linearized equations give the dispersion equation (for $\delta\rho\sim e^{-i\omega t+i k x}$ etc.)
\be
	\omega^{2}=c^{2} k^{2}+\frac{1}{2}A_{0} k^{4}-2i\nu_{0}\omega k^{2},
 \la{disp-eq}
\ee
where $c^{2}=\partial_{\rho}P|_{\rho_{0}}$ is a linear sound velocity and $\nu_{0}=\frac{\eta_{B0}}{2\rho_{0}}$ is proportional to the kinematic bulk viscosity at $\rho=\rho_{0}$.  Solving (\ref{disp-eq}) and expanding up to $k^{3}$ we arrive to
\be
	\omega(k) \approx \pm ck\left(1+\frac{A_{0}-2\nu_{0}^{2}}{4c^{2}}k^{2}\right) - i\nu_{0}k^{2}.
\ee
For the wave propagating to the right we have
\be
	\omega(k) \approx ck - i\nu_{0}k^{2}+\frac{A_{0}}{4c}k^{3} .
  \la{disp-eq-right}
\ee
Here we kept only leading terms representing different effects (dissipation $\nu_{0}$) and dispersion ($A_{0}$). We consider the limit of small dissipation and dispersion and, therefore, neglected the correction to the dispersion quadratic in $\nu_{0}$ assuming it smaller than the corresponding $A_{0}$ contribution. We keep the term $A_{0}k^{3}$ which can be comparable in magnitude to the term $i\nu_{0}k^{2}$ at small (but finite) $k$ if the viscosity $\nu_{0}$ is very small.
The dispersion (\ref{disp-eq-right}) can be reproduced by the following linear equation
\bea
	u_{t}+cu_{x}-\alpha u_{xxx} =\nu_{0}u_{xx}\, ,
 \la{kdvb-lin}
\eea
where $\alpha=A_{0}/(4c)$.
Here one can substitute either $\delta\rho$ or $\delta v$ instead of $u$ as $\delta \rho$ and $\delta v$ are related linearly by (\ref{cont-lin}). To obtain an effective equation for the left moving linear wave one should just change $x\to -x$ in (\ref{kdvb-lin}).

In the linear approximation an arbitrary initial profile of density and velocity is split into right and left moving waves. These waves move with velocities $\pm c$ respectively and slowly disperse and decay. If the amplitude of the initial profile is small but finite one should add nonlinear corrections to chiral (right and left) equations (\ref{kdvb-lin}). In addition to this, the nonlinear terms also couple the equations for left and right moving waves. However, as the right and left profiles pass each other with finite velocity (with the relative velocity $2c$) they interact only for a short time.  In the limit of weak nonlinearity the coupling between equations is small and will not significantly change solutions. Therefore, it can be neglected\cite{novikov} and most important nonlinear corrections should enter the chiral wave equation (\ref{kdvb-lin}) itself.

The goal of the next section is to justify the above arguments and to derive (\ref{kdvb-lin}) together with the corresponding nonlinear corrections using the well-known reductive perturbation method\cite{su,jeffery,wh,carr}.

\section{KdVB equation via Reductive Perturbation method}
 \la{sec:redpert}

The linear equation (\ref{kdvb-lin}) suggests that in the long wave approximation $k\to 0$ the dissipative $\nu_0$ term always wins over the dispersive $\alpha$ term. This is true unless the viscosity coefficient $\nu_{0}$ is small (or zero) for some reason and we are interested in the regime of small but finite $k$. Then the comparison of the first and third terms on the l.h.s. of (\ref{kdvb-lin}) suggests that we have to treat all derivatives as small with their relative values determined by the  scaling $\partial_{t}=O(\epsilon^{3})$ and $\partial_{x}=O(\epsilon)$ where $\epsilon$ is some auxiliary ``counting'' parameter treated as small parameter. This scaling was introduced in Ref.~\onlinecite{GardnerMorikawa-1960} in the derivation of the Korteweg--de Vries equation for water waves. The only modification we make here is that we consider the viscosity coefficient $\nu_{0}=O(\epsilon)$ which would allow us to have the r.h.s. of (\ref{kdvb-lin}) of the same order as $u_{t}$ and $u_{xxx}$. The aim of this section is to include a weak nonlinearity into this scaling. This goal is achieved by using the so-called reductive perturbation method (or a proper power counting scheme in modern language).

\subsection{Korteweg--de Vries--Burgers equation (KdVB)}
\label{kdvb_derivation}

Let us introduce the following scaling scheme
\begin{eqnarray}
	\delta\rho& \to &\epsilon^2 f(\epsilon \xi,\epsilon^{3}t),
 \label{kdvb_rho} \\
 	\delta v&\to &\epsilon^2 g(\epsilon \xi,\epsilon^{3}t),
 \label{kdvb_v} \\
 	\eta_{B} &\to & \epsilon \eta_{B},
 \label{kdvb_nu}
\end{eqnarray}
where
\be
	\xi = x-ct-x_{0}.
\ee
The scaling laws $\epsilon \xi$, $\epsilon^{3}t$ and $\epsilon \eta_{B}$ have been already motivated on the basis of linear equation (\ref{kdvb-lin}) in the beginning of this section. The $\epsilon^{2}$ scaling of the amplitudes can be obtained by comparing the relative values of, say, the term $u_{t}\sim \epsilon^{3} u$ of (\ref{kdvb-lin}) with a foreshadowed nonlinear correction $uu_{x}\sim \epsilon u^{2}$. Then according to a general reductive perturbation method the scaling scheme (\ref{kdvb_rho},\ref{kdvb_nu}) should be supplemented by an appropriate perturbation scheme
\bea
	f &=& f^{(0)}+\epsilon^{2}f^{(1)}+\ldots \, ,
 \la{pertf} \\
 	g &=& g^{(0)}+\epsilon^{2}g^{(1)}+\ldots \, .
 \la{pertg}
\eea
Substituting $\rho=\rho_{0}+\epsilon^{2} f(\epsilon \xi, \epsilon^{3}t) $ and $v=\epsilon^{2}g(\epsilon \xi, \epsilon^{3}t)$ with (\ref{pertf},\ref{pertg},\ref{kdvb_nu}) into the system (\ref{cont},\ref{euler}) we obtain a system of coupled equations which can be analyzed in each order of $\epsilon$ separately. The first non-vanishing order is $O(\epsilon^{3})$ gives the following two equations
\bea
	\frac{g^{(0)}}{c} &=& \frac{f^{(0)}}{\rho_{0}},
 \la{fg0}\\
 	f^{(0)} w_{0}' &=& c g^{(0)}.
 \nonumber
\eea
A compatibility of these equations determines sound velocity $c$ giving the well-known thermodynamic relation
\be
	c^{2}=\rho_{0}w_{0}'.
 \la{c-gen}
\ee
The equation (\ref{fg0}) gives a linear relation between $f^{(0)}$ and $g^{(0)}$.

The next non-vanishing order is $O(\epsilon^{5})$. It gives
\begin{eqnarray}
	\partial_{t}f^{(0)}+\partial_{\xi}(f^{(0)}g^{(0)}) &=& \partial_{\xi}(cf^{(1)}-\rho_{0}g^{(1)}),
 \nonumber \\
 	\partial_{t}g^{(0)}+\partial_{\xi}\left(\frac{{g^{(0)}}^{2}}{2} +\frac{w_{0}''}{2}{f^{(0)}}^{2}\right)
	&-& \frac{A_{0}}{2\rho_{0}}\partial_{\xi}^{3}f^{(0)}
 \nonumber \\
 	-\frac{\eta_{B0}}{\rho_{0}}\partial_{\xi}^{2}g^{(0)} &=&
	\partial_{\xi}(cg^{(1)}-w_{0}'f^{(1)}).
 \nonumber 
\end{eqnarray}
We use the relation (\ref{fg0}) to exclude $g^{(0)}$ from these equations. Then
the difference of these two equations gives a closed equation for $u=f^{(0)}$:
\bea
u_{t}+\zeta u u_{\xi}-\alpha u_{\xi\xi\xi}=\nu u_{\xi\xi}\qquad\mbox{\textbf{(KdVB)}}
 \la{kdvb}
\eea

with
\bea
	\zeta &=& \frac{1}{2}\left(\frac{3c}{2\rho_{0}}+\frac{w^{\prime\prime}_{0}\rho_{0}}{2c}\right)
	=\frac{1}{2}\left(\frac{c}{\rho_{0}}+\frac{\partial c}{\partial \rho_{0}}\right),
 \la{zeta-gen} \\
 	\alpha &=& \frac{A_{0}}{4c},
 \la{alpha-gen}\\
 	\nu &=& \nu_{0}=\frac{\eta_{B0}}{2\rho_{0}}.
 \la{nu-gen}
\eea
A similar equation for $g^{(0)}$ can be obtained from (\ref{kdvb}) using (\ref{fg0}). The equation (\ref{kdvb}) is known as the Korteweg--de Vries--Burgers equation (KdVB). It describes the correct far-field behavior of solutions of (\ref{cont},\ref{euler}) in the limit of weak nonlinearity, dispersion and dissipation. A linearization of (\ref{kdvb}) gives (\ref{kdvb-lin}) as it was expected.

We note here that in the conventional form of KdVB equation (and KdV equation, see below) the $\alpha$ term is usually written with plus sign. There is, however, a simple transformation
\begin{equation}
	u(x)\to -u(-x)
 \la{ama}
\end{equation}
which maps the solutions of equations with opposite signs of $\alpha$ to each other. Using this transformation we can borrow any solutions of conventional KdVB and KdV equations to find the corresponding solutions of (\ref{kdvb}).

\subsection{Scales and phase diagram of KdVB}
\label{pdKdVB}

KdVB has been extensively studied (see e.g., Refs. \onlinecite{kdvb1,kdvb2,kdvb3,kdvb4,bona}. It is a non-integrable equation and only few traveling wave solutions of KdVB are known analytically\cite{kdvb1}. KdVB has a reach dynamics which includes an intricate interplay of nonlinearity, dispersion and dissipation. An example of numerical analysis of KdVB can be found in Ref.~\onlinecite{kdvb4}.


The effective KdVB equation (\ref{kdvb}) with (\ref{zeta-gen},\ref{alpha-gen},\ref{nu-gen}) describes the universal limit of weak nonlinearity, dispersion and dissipation of one-dimensional solutions for a large class of hydrodynamic systems. Depending on initial conditions different terms of (\ref{kdvb}) play different role at different times of evolution. The goal of this section is to summarize different regimes of hydrodynamic behavior and to show how to make simple estimates for onsets of these regimes.

To understand qualitatively the role of different effects let us consider an initial profile $u(x,t=0)$ which is characterized just by two scales: a typical width of the profile $W$ and a typical amplitude of the initial profile $U$. One can think of this initial profile as of the step-like function $u(\xi)$ shown in the insert of Fig.~\ref{estimate} or as of the single lump of density (velocity).
We use the parameters $U$ and $W$ together with the values of coefficients $\zeta$, $\alpha$, $\nu$ to form characteristic times corresponding to the different terms of KdVB. We characterize the strength of  nonlinearity, dispersion  and dissipation by the following inverse times (characteristic frequencies)
\bea
	\Omega_{\zeta}=\zeta UW^{-1}, \quad \Omega_{\alpha}=\alpha W^{-3},
	\quad \Omega_{\nu}=\nu W^{-2},
 \la{scales}
\eea
respectively. The relative values of these frequencies define the regime of the evolution of the initial profile described by KdVB. Different regimes are summarized in the Figure~\ref{estimate}. It is also convenient to introduce the spatial scales
\bea
	W_{\zeta\nu}=\frac{\nu}{\zeta U}, \quad W_{\zeta\alpha}=\sqrt{\frac{\alpha}{\zeta U}},
	\quad W_{\nu\alpha}=\frac{\alpha}{\nu}.
 \la{Wscales}
\eea
These scales are defined so that, e.g., when $W\sim W_{\zeta\nu}$ the $\zeta$- and $\nu$-terms of KdVB are of the same order $\Omega_{\zeta}\sim \Omega_{\nu}$.

The center of the triangle in Figure~\ref{estimate} represents an initial profile for which all three scales are of the same order, i.e. nonlinearity, dispersion and dissipation are equally important in the beginning of the evolution. Generally, different initial profiles correspond to different points of the triangle so that the corresponding parameters grow from zero at the respective side to the infinity in the respective vertex of the triangle.

In the absence of nonlinearity (the bottom part of the triangle, $\Omega_{\zeta}\ll \Omega_{\alpha,\nu}$) the dynamics is approximately linear and is described by (\ref{kdvb-lin}). It is either diffusive (left part of the triangle, $\Omega_{\alpha}\ll\Omega_{\nu}$) or initially dispersive (right part, $\Omega_{\nu}\ll \Omega_{\alpha}$). The dispersive evolution even if dominant in the beginning of the evolution $\Omega_{\alpha}\gg\Omega_{\nu}$ leads to the growth of the width of the initial profile and the decrease of gradient terms. The dispersion term will become of the same order as the diffusion one at the time
\bea
	t_{\alpha\nu} \sim \Omega_{\alpha}\Omega_{\nu}^{-2}=\alpha\nu^{-2}W.
 \la{tnu}
\eea
At that time the width of the profile becomes of the order of $W_{\nu\alpha}$ (\ref{Wscales}) and after that moment the dispersion becomes subdominant to the diffusion.

The evolution is much more interesting and complicated when the nonlinear term dominates for an initial profile. In this case the evolution is initially described by the Riemann-Hopf (or inviscid Burgers) equation
\be
	u_{t}+\zeta uu_{\xi} = 0, \qquad\mbox{\textbf{(Riemann-Hopf})}
 \la{RHeq}
\ee
which can be easily solved for any initial profile $u(x,t=0)=f(x)$ giving $u(x,t)$ implicitly, as a solution of $u=f(x-\zeta u t)$, where the unknown $u$ enters both left and right hand sides.
For our typical initial $UW$-profile this solution is well defined at small times becoming multiply-valued for times after
\be
	t_{c}\sim \Omega_{\zeta}^{-1}=W/(\zeta U).
 \la{tc}
\ee
The time $t_{c}$ is known as the time of ``gradient catastrophe'' and is defined as the time at which $u_{x}$ becomes infinite. The classical problem (\ref{RHeq}) is ill-posed at larger times and has to be regularized by higher gradient corrections which in our case are either dispersive ($\alpha$-term of (\ref{kdvb})) or diffusive ($\nu$-term). According to the relative strength of subdominant diffusive and dispersive terms in the regime of strong nonlinearity (the upper part of the triangle phase diagram) one has either formation of classical dissipative shock waves (left part) or dispersive shock waves (right part). We discuss these shock wave regimes separately in Sec.~\ref{sec:dsw}. Before going to this discussion we consider important limits of KdVB equation (\ref{kdvb}) corresponding to the left and right sides of the triangle phase diagram of Fig.~\ref{estimate}.

\subsection{Dissipative limit: Burgers equation}
\label{Burgers_limit}

If the dispersion does not play a significant role in the evolution ($\Omega_{\alpha}\ll\Omega_{\nu}$) one can neglect $\alpha$-term in (\ref{kdvb}) and arrive to the well-known Burgers equation
\begin{equation}
	u_{t}+\zeta u u_{\xi}
	=\nu u_{\xi\xi}\qquad\mbox{\textbf{(Burgers)}}.
 \label{burgers}
\end{equation}
The regime described by (\ref{burgers}) corresponds to the left side of the triangle phase diagram of Fig.~\ref{estimate}. The different points of the left side the triangle correspond to different relative values of nonlinear and dissipative terms of (\ref{burgers}).

We note here that one can arrive to the Burgers equation (\ref{burgers}) more formally starting directly from (\ref{cont},\ref{euler}). Instead of (\ref{kdvb_nu}) one should take $\eta_{B} = O(\epsilon^{0})$. Then the scaling scheme (\ref{kdvb_rho},\ref{kdvb_v}) is not consistent and should be replaced by the Burgers scheme\cite{su}
\begin{eqnarray}
 \label{bur_rho}
	\delta\rho&=&\epsilon f(\epsilon x,\epsilon^{2}t),
 \\
 	v&=&\epsilon g(\epsilon x,\epsilon^{2}t)
 \label{bur_v}
\end{eqnarray}
with the corresponding perturbative expansion
\begin{eqnarray}
 \label{bur_per_f}
	f&=&f^{(0)}+\epsilon f^{(1)},
 \\
 	g&=&g^{(0)}+\epsilon g^{(1)}
\label{bur_per_g}
\end{eqnarray}
instead of (\ref{pertf},\ref{pertg}).
Expanding in $\epsilon$ we arrive to the relation (\ref{fg0}) and to the Burgers equation (\ref{burgers}) as a closed equation for $u=f^{(0)}$ with parameters given by (\ref{zeta-gen},\ref{nu-gen}).

The Burgers equation (\ref{burgers}) is often used as a model describing classical dissipative shock waves which result from the interplay between nonlinearity and dissipation.\cite{AblowitzEtAl} It can be analytically solved via Cole-Hopf transformation $u=-\frac{2\nu}{\zeta}\partial_{\xi}\log\phi$ which results in a diffusion equation $\phi_{t}=\nu\phi_{xx}$. \cite{ch,ch1}


%

There is a simple exact solution of (\ref{burgers}) given by a step-like profile traveling with a constant velocity $V$ to the right
\be
	u(\xi,t) = \frac{U}{2}\Big(1-\tanh\left[W^{-1}(\xi-Vt)\right]\Big).
 \la{burgers-shock-wave}
\ee
Here the parameters of the step are related to the velocity $V$ as
\bea
 	W &=& 2\nu/V=4W_{\zeta\nu}\, ,
 \la{W-shock} \\
	U &=& 2V/\zeta\, .
 \la{U-shock}
\eea
The width of the solution is given by (\ref{W-shock}), i.e., essentially by the scale $W_{\zeta\nu}$.
In the limit $\nu\to 0$ this width goes to zero  and the solution (\ref{burgers-shock-wave}) describes a discontinuous shock front. The solution (\ref{burgers-shock-wave}) is defined when $\Delta u=u_{\xi\to-\infty}-u_{\xi\to+\infty}=U>0$.



\subsection{The limit of no dissipation: KdV equation}
\label{kdv_limit}

In the limit when dissipation is absent or very small at a given wavelength the $\nu$ term of KdVB (\ref{kdvb}) can be dropped and we arrive to the celebrated Korteweg--de Vries equation (KdV)
\begin{equation}
	u_{t}+\zeta u u_{\xi}-\alpha u_{\xi\xi\xi}
	=0\qquad{\mbox{\textbf{(KdV)}}}
 \label{kdv}
\end{equation}
with parameters determined by (\ref{zeta-gen},\ref{alpha-gen}). It is a purely dispersive equation.  It is integrable\cite{miura,kdv-zakharov} and possesses infinitely many conserved quantities. The first three of these integrals are given explicitly by
\bea
	I_{0} &=& \int d\xi \, u \, ,
 \la{I0}\\
 	I_{1} &=& \int d\xi \, \frac{u^{2}}{2}\, ,
 \la{I1}\\
 	I_{2} &=& \int d\xi \, \left[\zeta \frac{u^{3}}{3} +\alpha u_{\xi}^{2}\right]\, .
 \la{I2}
\eea
The integrals $I_{0,1,2}$ are related to the total number of particles, total momentum and total energy of the system (\ref{cont},\ref{euler}) as follows
\bea
	N-N_{0} &=& I_{0},
 \la{NI0}\\
 	P &=& \frac{2c}{\rho_{0}}I_{1}+cI_{0},
 \la{PI1}\\
 	E-E_{0} &=& \frac{c}{\rho_{0}}I_{2}+\frac{2c^{2}}{\rho_{0}}I_{1}+w_{0}I_{0}.
 \la{EI2}
\eea
The higher integrals of motion of KdV are related to more complicated symmetries of the problem and are usually destroyed by small corrections to KdV which destroy integrability. We do not need their exact form in the following discussion. The conserved quantities (\ref{NI0},\ref{PI1}) and (\ref{EI2}) on the other hand are related to fundamental space-time-gauge symmetries and, therefore, play an important role in the hydrodynamic approach.

The KdV has a solitary wave solution (soliton) moving to the left with velocity $V$\footnote{The corresponding solution of original system (\ref{cont},\ref{euler}) moves to the right with velocity $c-V$.}
\be
	u(\xi,t) = -U \cosh^{-2}\Big(W^{-1}(\xi+Vt)\Big).
\label{kdv-soliton}
\ee
This solution corresponds to a local depletion of the particle density (minus sign in (\ref{kdv-soliton})) and is known as the dark soliton. The width $W$ of the soliton and the amplitude $U$ of the depletion in (\ref{kdv-soliton}) are defined by its velocity $V$ as
\bea
	W &=& (4\alpha/V)^{1/2}=\sqrt{12}\, W_{\zeta\alpha}\, ,
 \la{W-KdV}\\
 	U &=& 3V/\zeta \, .
 \la{U-KdV}
\eea
While the parameters of the soliton (\ref{kdv-soliton}) are given exactly by (\ref{W-KdV},\ref{U-KdV}) they could also be estimated from the condition that the nonlinearity and dispersion scales of the soliton solution should be of the same order $\Omega_{\alpha}\sim \Omega_{\zeta}$ (exactly we have: $\Omega_{\zeta}=12\, \Omega_{\alpha}$). \footnote{The velocity of the soliton $V$ can also be estimated by equating the first term of (\ref{kdvb}) to the second $U/T \sim \zeta U^{2}W^{-1}$, where $T$ is the typical time so that the typical velocity is $W/T \sim \zeta U$. Indeed, from (\ref{U-KdV}) we have exactly $V=\zeta U/3$.}

The solution (\ref{kdv-soliton}) corresponds to the special point on the right side of the triangle phase diagram in Fig.~\ref{estimate}. These solutions (as the effective KdV itself) are to be trusted only when the relative depletion $U/\rho_{0}\sim V/c$ is small, i.e., $V\ll c$.

The total number of depleted particles in the soliton solution (\ref{kdv-soliton},\ref{W-KdV},\ref{U-KdV}) is given by the zeroth integral of motion (\ref{I0})
\be
	n=\int u(\xi,t)\,d\xi = -2UW=-\frac{12}{\zeta}\sqrt{\alpha V}\, .
 \la{n-KdV}
\ee
The values of other integrals of motion on the KdV soliton solution (\ref{kdv-soliton})
are
\bea
	I_{1} &=& \frac{2}{3}U^{2}W=12 \frac{V \sqrt{\alpha V}}{\zeta^2}
	=-\frac{V}{\zeta}n\, ,
 \\
 	I_{2} &=& -\frac{4}{15}\zeta U^{3} W =-\frac{72}{5}\frac{V^{2}}{\zeta^{2}}\sqrt{\alpha V}
	=\frac{6}{5}\frac{V^{2}}{\zeta} n\, .
\eea
Using above expressions and  (\ref{NI0},\ref{PI1},\ref{EI2}) we obtain the dispersion of the soliton
\bea
	E(P)-E_{0}-nw_{0}=c\left(P-cn\right)+\frac{\left(P-cn\right)^{2}}{2m^{*}}
 \la{soldisp}
\eea
with the ``effective mass'' $m^{*}$ given by
\bea
	m^{*}=\frac{5cn}{3\zeta\rho_{0}}
\label{mstar}
\eea
The form (\ref{soldisp}) is similar to the dispersion of particles with quadratic spectrum except that the effective mass (\ref{mstar}) is velocity (and momentum) dependent (see eq.~\ref{n-KdV}).

Another important exact solution of KdV is the periodic traveling wave solution given by
\be
	u(\xi,t) = -U \mbox{cn}^{2}
	\left(W^{-1}(\xi+Vt)\Big|m\right).
 \label{kdv-per}
\ee
In Eq.~\ref{kdv-per}, $\mbox{cn}(y|m)$ is the Jacobi elliptic function of modulus $m$ $(0 < m < 1)$. The modulus defines the period of the solution in $\xi$ (the period $L$ itself is $L=W\,F(\pi/2,m)$). In the limit $m\rightarrow 1$ the period $L\to \infty$ and Jacobi elliptic function reduces to a periodic array of well-separated solitons (\ref{kdv-soliton}).

So far we discussed very special solutions of KdV corresponding to the ``Solitons'' point in Fig.~\ref{estimate}. What will happen to the initial profile with typical dimensions such that the nonlinear term is dominant $\Omega_{\zeta} \gg\Omega_{\alpha}$? In this case the profile will
initially evolve according to the Riemann-Hopf equation (\ref{RHeq}). However, at the gradient catastrophe time (\ref{tc}) large gradients will develop and the dispersive term will become of the order of the nonlinear term. Details of the evolution after that point do depend on a particular shape of the initial profile. However, generically, the modulated periodic solutions of KdV will be generated providing oscillating (a.k.a.~dispersive) shock fronts \cite{GurevichPitaevskii-1973}. This phenomenon is referred to as Dispersive Shock Waves (DSW) in contrast to conventional or dissipative shock waves which occur in equations of Burgers type. Without dissipation the steep profiles described by KdV generate oscillations and eventually decay into trains of well-separated solitons.

\section{Shock waves in KdVB}
\la{sec:dsw}

\begin{figure}
\begin{center}
\includegraphics[scale=.50,angle=0]{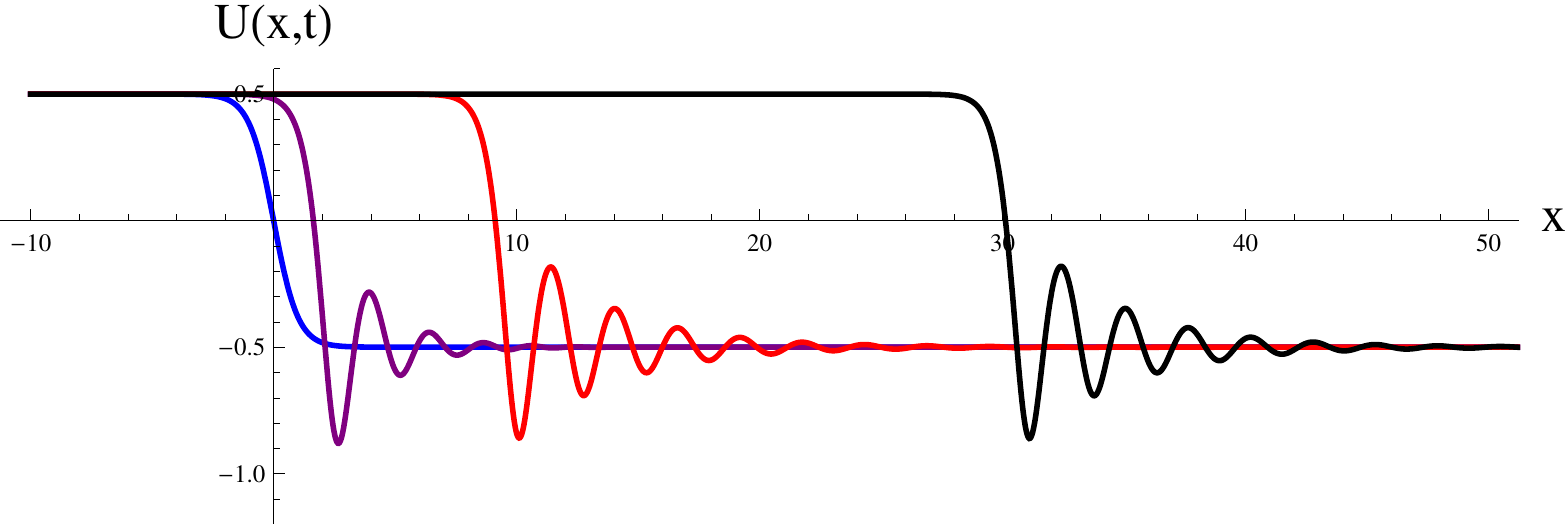}
\caption{The evolution of the initial step-like profile governed by the KdVB equation (\ref{kdvb}) in the dispersive shock wave regime.}
\label{fig_kdvb_dsw_compact}
\end{center}
\end{figure}

So far we have discussed the dynamical regimes of KdVB corresponding to the sides of a triangle phase diagram Fig.~\ref{estimate}. These regimes are realized when one of three scales of KdVB (\ref{scales}) can be neglected compared to two others. If all three scales are significant the different ordering of these scales roughly divide the phase diagram of Fig.~\ref{estimate} into six parts. Up to this point we treated three scales as equal but to understand the KdVB dynamics of a typical profile even better one should realize that the the nonlinearity, dissipation and dispersion enter the game in a very asymmetric way. Indeed, we have already seen that in the case when the nonlinearity dominates an initial evolution ($\Omega_{\zeta}$ is the largest scale) the evolution itself will result in large gradients in finite time which is of the order of the gradient catastrophe time (\ref{tc}). At that point either dissipation or dispersion will become of the order of nonlinearity and our discussion of further evolution should start from the different part of the triangle phase diagram \footnote{One can even draw RG-flow-type diagram where the point representing $UW$ profile moves within the triangle in a real time. However, the profile seizes to be a simple $UW$ profile during this ``flow'' and one should be more precise describing different parts of the composite profile characterized by different scales.}  Another example of this type is given by the linearized KdVB (\ref{kdvb-lin}) with dominating dispersive term ($\Omega_{\alpha}\gg\Omega_{\nu}$). The corresponding initial profile will disperse almost non-dissipatively up to the time of the order of
$t_{\alpha\nu}$ (\ref{tnu}) at which point the dissipation term will become of the order of the dispersion term and the further evolution will be diffusion-like. In contrast to these examples, if the dissipation term is dominant in the beginning it remains dominant and the evolution is diffusion-like at all times.

In the following we apply the understanding of the asymmetry of different KdVB terms to describe two distinct shock wave regimes of KdVB. We consider an initial profile of the step form shown in the insert of Fig.~\ref{estimate}. Shock waves appear in the regime when nonlinearity is the most important scale in the problem, e.g., if the width of the initial profile $W$ is sufficiently bigger than $W_{\zeta\nu}$ and $W_{\zeta\alpha}$  (\ref{Wscales}). Then the nature of shock waves depends on the relative strength of dispersive and dissipative terms.

The goal of this section is to describe the major features of classical and dispersive shock waves and show how to estimate typical scales of these shock wave solutions.

\subsection{Classical shock waves (CSW)}
\label{csw}

To have a clear separation of scales we consider the following relation between the width of the initial step-like profile and scales of KdVB (\ref{Wscales})
\bea
	W \gg W_{\zeta\nu} \gg W_{\zeta\alpha},\,W_{\nu\alpha}.
 \la{csw-regime}
\eea
which means $\Omega_{\zeta}\gg\Omega_{\nu}\gg\sqrt{\Omega_{\alpha}\Omega_{\zeta}}$.
In this case the dominant term of KdVB for the initial profile is the nonlinear $\zeta$-term and the evolution of this profile is determined by the competition between dominant nonlinear and subdominant dissipative $\nu$-term.

Up to the time of the order of $t_{c}$ (\ref{tc}) the nonlinear term will dominate the evolution which will be described by the Riemann-Hopf equation (\ref{RHeq}). Then the decreasing width of the profile (the width of the wave front) will reach the scale $W_{\zeta\nu}$ (\ref{scales}) and an approximate balance between nonlinearity and dissipation will be achieved. The profile will become stationary and will have a width of the order of
\begin{equation}
	W_{CSW}\sim W_{\zeta\nu} 
 \la{W-CSW}
\end{equation}
The quasi-stationary profile can be roughly described by the solution (\ref{burgers-shock-wave}). The ratio of the dispersion scale to the dissipative scale for this profile is given by $\Omega_{\alpha}/\Omega_{\nu}=W_{\nu\alpha}/W$ and will remain very small for all times due to (\ref{csw-regime}). Therefore, we expect that the dispersive effects are not important in the regime (\ref{csw-regime}) and might result only in small oscillations on top of the stationary classical shock wave solution (\ref{burgers-shock-wave}).

\subsection{Dispersive shock waves (DSW)}
\label{dsw-sub}

The case of the initial step-like profile characterized by the width
\bea
	W,\, W_{\nu\alpha} \gg W_{\zeta\alpha} \gg W_{\zeta\nu}
 \la{dsw-regime}
\eea
is, probably, the most interesting.
The inequalities (\ref{dsw-regime}) mean that the evolution of the initial profile is defined in the beginning by an interplay between the dominant nonlinear and subdominant dispersion terms with dissipation playing some role only at very large times. As both $\nu$- and $\alpha$-terms are subdominant to the nonlinearity the scale $W_{\nu\alpha}$ does not play any role in the beginning of the evolution and its relation to the initial profile width $W$ is not very important. We note also that the second of the inequalities (\ref{dsw-regime}) is a consequence of the first one and the smallest scale $W_{\zeta\nu}$ does not play a major role in this regime.

As in the case of classical shock waves for $t<t_{c}$ the evolution is governed by the Riemann-Hopf equation (\ref{RHeq}) and the width of the step decreases reaching $W_{\zeta\alpha}$ in time of the order of $t_{c}$ (\ref{tc}). At this time oscillations develop at the trailing edge of the step profile\footnote{For more conventional form of KdVB with plus sign in front of the $\alpha$ term, the oscillations develop at the leading edge of the shock front which is in agreement with (\ref{ama}).} These oscillations grow in amplitude with the largest amplitude becoming of the order of the size of the step $U$. At this point the typical wavelength of oscillations is given by the scale $W_{\zeta\alpha}$ from (\ref{Wscales}). The number of oscillations and the spatial extent of the oscillating part of the shock front continue to grow. In the absence of dissipation (KdV) this growth continues forever with oscillations evolving into the train of well-separated solitons \cite{GurevichPitaevskii-1973}.

In the case of non-vanishing dissipation the front of DSW becomes stationary at large times. The width of the stationary profile can be estimated in the following simple way. As the width of the oscillating shock profile is determined by the dissipation we evaluate the magnitude of the dissipative term $\Omega_{\nu}$ (\ref{scales}) at the gradients developed due to oscillations, i.e., $\Omega_{\nu}\Big|_{W_{\zeta\alpha}}\sim \frac{\nu}{W_{\zeta\alpha}^{2}}=\frac{\nu}{\alpha}\zeta U$. On a stationary profile this scale should be of the order of the nonlinear scale evaluated at  the overall width of the profile $W_{DSW}$, i.e. $\Omega_{\zeta}\Big|_{W_{DSW}}\sim \frac{\zeta U}{W_{DSW}}$. Equating these scales we obtain an estimate
\begin{equation}
	W_{DSW} \sim \frac{W_{\zeta\alpha}^{2}}{W_{\zeta\nu}}=\frac{\alpha}{\nu}=W_{\nu\alpha}.
 \la{W-DSW}
\end{equation}
The formation of the stationary shock front profile of DSW takes time of the order of $t_{DSW}\sim W_{\nu\alpha}/(\zeta U)$ and the number of oscillations in the stationary profile can be estimated as
\bea
	N\sim \frac{W_{\nu\alpha}}{W_{\zeta\alpha}}
	\sim \left(\frac{\alpha\zeta U}{\nu^{2}}\right)^{1/2}\gg 1\,.
 \la{Nosc}
\eea
This condition can be considered as a necessary condition of observing oscillations in the stationary shock wave profile.

We have to remark here that the presented picture captures only major scales of the dispersive shock wave formation. For example, while we assumed for our estimates that the typical amplitude of oscillations is $U$, the oscillations at all amplitudes ranging from $0$ to $\sim U$ are present in the shock front. As a result even after the formation of the stationary shock wave profile for major scale $U$ the small amplitude oscillations keep propagating at the leading edge of the shock wave profile. The reader is referred to the seminal Ref.~\onlinecite{GurevichPitaevskii-1987} for details. In Ref.~\onlinecite{GurevichPitaevskii-1987} Gurevich and Pitaevskii used the large parameter separating the scales $W_{\nu\alpha}/W_{\zeta\alpha}\gg 1$ to describe the formation of the DSW profile of KdVB analytically using the Whitham modulation theory. They described the oscillating part of the profile by a modulated periodic solution of KdV (\ref{kdv-per}) (see also Ref. \onlinecite{AblowitzEtAl} for a recent discussion of DSW in KdV and cold atom dynamics).

We conclude that in striking contrast to the classical shock wave the dispersive shock wave profile has an internal structure -- oscillations with the typical wavelength $W_{\zeta\alpha}$ and that while the width of the conventional shock wave is proportional to $\nu$ (\ref{W-CSW}), the overall width of DSW is proportional to $\nu^{-1}$ (\ref{W-DSW}).

The main steps of the formation of DSW are illustrated in Fig.~\ref{fig_kdvb_dsw_compact}. Although the KdVB equation itself is considered in this work only as an approximation to more precise hydrodynamics (\ref{cont},\ref{euler}) it is important that it gives a very good qualitative and often even quantitative understanding of the latter. Below, in section \ref{UFG}, using the numerical method of smoothed particle hydrodynamics, we present the results of numeric solutions of hydrodynamic equations written for a cold Unitary Fermi gas (see Fig.~\ref{fig_comp}). The similarity between these solutions and the dispersive shock waves of KdVB described in this section are evident. Using the results given in this section we will make estimates (in Sec \ref{UFG}) for dispersive shock wave scales in a Unitary Fermi gas system for gas parameters similar to the recent experiment\cite{jjka}.

\begin{table*}[tp] 
\caption {Hydrodynamic descriptions for some cold atom systems. Mapping of the parameters in hydrodynamic equations (\ref{cont},\ref{euler}) to the coefficients of the effective KdVB equation.
}
\label{table_values}%
\begin{tabular}{|c|c|c|c|c|c|c|c|}
\hline
\textbf{Cold Atomic Model} & $w(\rho)$ &
$\frac{m^{2}}{\hbar^{2}}A(\rho)$ & $\frac{m}{\hbar}\eta_{B}(\rho)$ &
$c^{2}=\rho_{0}w_{0}'$ & $\zeta=\frac{3c}{4\rho_{0}}$ &
$\alpha=\frac{A_{0}}{4c}$ &
$\nu=\frac{m}{\hbar}\eta_{B0}/\rho_{0} $ \tabularnewline
\textbf{\textit{Chiral differential equation}}  & & & & &
$+\frac{w^{\prime\prime}\rho_{0}}{4c}$& & \tabularnewline
\hline
\hline
\hline
Dense Lieb-Liniger Gas& $g\rho/m$ & $1/2 $ & $0^{a}$ &
$\frac{\hbar^2\rho_{0}^{2}}{m^2}\gamma$ &
$\frac{\hbar}{m}\frac{3}{4}\sqrt{\gamma}$
&$\frac{\hbar}{m}\frac{1}{8\rho_{0}\sqrt{\gamma}}$ & $0$
\tabularnewline
\textit{KdV} (Eq. \ref{kdv})  & & & & & & & \tabularnewline
\hline
Tonk's Gas& $\frac{\pi^{2}\hbar^{2}\rho^{2}}{2 m^{2}}$ & $0$ & $0^{a}$
& $\frac{\hbar^2\pi^{2}\rho_{0}^{2}}{m^2}$ & $\frac{\hbar}{m}\pi$ & 0
&  $0$ \tabularnewline
\textit{Riemann-Hopf} (Eq. \ref{RHeq}) & & & & & & & \tabularnewline
\hline
Plane-Wave Unitary Gas &
$(1+\beta)\frac{\hbar^{2}}{2m^{2}}(3\pi^{2}\rho)^{\frac{2}{3}}$ &
$1/2$ & $\sim \rho^{b}$ & $\frac{2}{3}w_{0}$ &
$\frac{2}{3}\frac{c}{\rho_{0}}$ & $\frac{\hbar^{2}}{8m^{2}c}$ &  $\sim
1$ \tabularnewline
\textit{KdVB} (Eq. \ref{kdvb}) & & & & & & &\tabularnewline
\hline
Quasi-1D Unitary Gas&
$\frac{\hbar^{2}}{2m^{2}l_{\perp}^{2}}(1+\beta)\left[\frac{15\pi}{2}\frac{\rho
l_{\perp}}{1+\beta}\right]^{\frac{2}{5}}$ & $9/20$ &  $\sim \rho^{b}$
& $\frac{2}{5}w_{0}$ & $\frac{3c}{5\rho_{0}}$ &
$\frac{9\hbar^{2}}{80m^{2}c}$& $\sim 1$\tabularnewline
\textit{KdVB} (Eq. \ref{kdvb}) & & & & & & & \tabularnewline
\hline
Calogero Gas &
$\frac{\hbar^{2}\lambda^{2}\pi^{2}\rho^{2}}{2m^{2}}
	+\frac{\hbar^{2}\lambda^{2}\pi\rho_{x}^{H}}{m^{2}}$ &
$\lambda^{2}/2$ &
$0$ &
$\frac{\hbar^2\pi^{2}\lambda^{2}\rho_{0}^{2}}{m^2}$ &
$\frac{\pi\hbar\lambda}{m}$ &
$\alpha_{BO}=\frac{\hbar\lambda}{2m}$ &
0\tabularnewline
\textit{Benjamin-Ono} (Eq. \ref{BO_eq_example}) &
 &
 &
 &
 &
 &
 &
 \tabularnewline
\hline
Quasi-1D Dipolar BEC &
$\frac{gA_{1D}\rho}{ml_{\perp}^{2}}+
\frac{gB}{m}\partial_{x}^{2}\rho^{U}$ &
$1/2 $ &
0&
$\frac{\hbar^{2}}{m^{2}l_{\perp}^{2}}\gamma A_{1D}$ &
$\frac{3}{4}\frac{c}{\rho_{0}}$&
$\alpha_{D}=\gamma
B\frac{\hbar^{2}}{2m^{2}c}$ &
0\tabularnewline
\textit{non-local KdV} (Eq. \ref{nkdv_eq_example}) &
&
&
&
&
&
$\alpha=\frac{\hbar^{2}}{8m^{2}c}$ &
\tabularnewline
\hline
\end{tabular}
$^{a}$In experiments, many atomic gases which to a good approximation
behave like Lieb-Liniger gas might still have mechanisms of
dissipation. The dissipation can be included phenomenologically by adding $\nu u_{\xi\xi}$ to the right hand side of effective differential equation which makes it KdVB (dense limit) or Burgers (Tonk's limit) equation.\\
$^{b}$See the discussion of viscosity in Unitary Fermi gas in Sec. VI B
for more details.
\end{table*}

\section{Interacting cold atoms}
\label{examples}

In this section we consider several one-dimensional models which recently attracted a lot of interest in connection with cold atom systems. These are the systems of bosons with contact interaction in both weak and strong coupling limits and the Fermi gas at unitarity. These models do respect the Galilean invariance and therefore can be described by the general form of (\ref{cont},\ref{euler}) in the hydrodynamic regime. The goal of this section is to relate the chemical potential $w(\rho)$ and the values of coefficients $A, \eta_{B}$ in (\ref{euler}) to the parameters (\ref{zeta-gen},\ref{alpha-gen},\ref{nu-gen}) of the effective KdVB (\ref{kdvb}) description. We delegate the analogous discussion of several models with long range interactions to the  Appendix \ref{long-range-app}. The results are summarized in the Table~\ref{table_values}.

\subsection{1D bosons with contact interaction}
\label{LLG}

One-dimensional bosons with contact interaction can be described by the Lieb-Liniger model\cite{LLM}
\begin{eqnarray}
 \label{LLE}
	H=-\frac{\hbar^{2}}{2m}\sum_{i=1}^{N}
	\frac{\partial^{2}}{\partial x_{i}^{2}}+g\sum_{i<j}\delta(x_{i}-x_{j}) \,.
\end{eqnarray}
The background density of bosons $\rho_{0}$ and the coupling constant $g$ define the
dimensionless coupling $\gamma$ as
\be
	\gamma = \frac{m}{\hbar^{2}\rho_{0}}g\, .
\label{gamma0}
\ee
The value of $\gamma$ determines whether the system is in a weak coupling $(\gamma\ll 1)$ or a strong coupling $(\gamma\gg 1)$ regime.

The model (\ref{LLE}) is integrable by the Bethe Ansatz method for all values of the coupling $\gamma$. In particular, a general form of $w(\rho)$ can be found implicitly through the solution of the Bethe Ansatz integral equations \cite{LLM}. For a general value of $\gamma$ this solution can be found only numerically. Here we discuss only the limits of weak and strong coupling in which analytical formulas can be obtained using the expansions in $\gamma$ and $1/\gamma$, respectively.

As the model (\ref{LLE}) is integrable it does not include any dissipation mechanism. However,  in more realistic modelings of experiments the dissipative effects might be significant. In simplest cases these effects can be incorporated into the effective one-dimensional hydrodynamic description by adding the Burgers terms $\nu u_{\xi\xi}$ with $\nu$ treated as a phenomenological parameter.

\subsubsection{Weak coupling (high density) limit $\gamma\ll 1$}

It is well-known that a collective description of (\ref{LLE}) in the high density limit $\gamma\ll1$  is given by the Gross-Pitaevski equation\cite{becrmp} (GPE), (see, e.g., Refs.~\onlinecite{becrmp,olshani})
\begin{eqnarray}
	i\hbar\partial_{t}\psi(x,t)
	&=&\Biggl\{-\frac{\hbar^2}{2m}\partial_{xx}+ g|\psi(x,t)|^{2}\Biggr\}\psi(x,t) \, .
 \la{GPE}
\end{eqnarray}
Using a ``hydrodynamic'' change of variables,
\begin{equation}
	\psi=\sqrt{\rho}\, e^{i\frac{m}{\hbar}\int_{0}^{x}v(x^\prime)\,dx^\prime}
\end{equation}
the GPE can be casted in the form (\ref{cont},\ref{euler}) with $w$, $A$ and $\eta_{B}$ given by
\bea
	w(\rho) &=& \frac{g}{m}\rho  \, ,
 \\
 	A &=& \frac{\hbar^{2}}{2m^{2}} \, ,
 \\
 	\eta_{B} &=& 0 \, .
\eea
Then from (\ref{c-gen},\ref{zeta-gen},\ref{alpha-gen}) we have
\bea
	c &=& \frac{\hbar\rho_{0}}{m}\sqrt{\gamma} \, ,
 \\
 	\zeta &=& \frac{\hbar}{m} \frac{3}{4}\sqrt{\gamma} \, ,
 \label{zetaGPE}
 \\
 	\alpha &=& \frac{\hbar}{m\rho_{0}}\frac{1}{8\sqrt{\gamma}} \, .
 \label{alphaGPE}
\eea
As mentioned earlier a small amount of dissipation that arises experimentally can be taken into account by introducing $\nu$ as a phenomenological parameter. Then, the dynamics of the model is described by the KdVB equation (\ref{kdvb}) with the values of $\zeta$ and $\alpha$ given by (\ref{zetaGPE},\ref{alphaGPE}), respectively, and with $\nu$ as a phenomenological parameter. In the limit of no dissipation one can describe this system by the KdV equation (i.e., KdVB with $\nu=0$).

The parameters of the soliton solution of an effective KdV equation in this case can be found from (\ref{W-KdV},\ref{U-KdV},\ref{n-KdV})
\bea
	W &=& \frac{1}{\rho_{0}}\sqrt{\frac{c}{2V\gamma}},
 \label{Wkdv_par} \\
 	U &=& 4\frac{V}{c}\rho_{0},
 \label{Ukdv_par} \\
 	n &=& -8\sqrt{\frac{V}{2c\gamma}},
 \label{nkdv_par}
\eea	
and the effective mass (\ref{mstar}) is
\bea
	m^{*}=\frac{20}{9}n.
\eea
This soliton solution of KdV coincides with the soliton solution of GPE equation (\ref{GPE}) in the limit of weak nonlinearity. The latter is known to correspond to a quasi-hole excitation of the quantum Lieb-Liniger model (\ref{LLE}) \cite{SklyaninQLL,KamenevQLL} The soliton solutions should be trusted only in the limit $|n|\gg 1$ and $V\ll c$. The first condition corresponds to the applicability of GPE while the second one allows us to replace GPE by KdV corresponding to a weakly nonlinear limit of GPE.

\subsubsection{Strong coupling (low density) limit $\gamma\gg 1$}

In the limit $\gamma\gg 1$ the  chemical potential is given by the following expansion in $1/\gamma$ \cite{LLM,guan,rigol}
\begin{equation}
	w(\rho)=\frac{\pi^{2}\hbar^2\rho^{2}}{2m^2}\left[1-\frac{4}{3\tilde\gamma}
	+\frac{20}{\tilde\gamma^{2}}
	+\frac{64\left(\frac{\pi^{2}}{15}-1\right)}{\tilde\gamma^{3}}+...\right]
\end{equation}
where $\tilde\gamma(\rho)$ is understood as a function of the local density $\rho$ (cf. Eq. \ref{gamma0})
\be
	\tilde\gamma(\rho) = \frac{m}{\hbar^{2}\rho}g.
 \label{gammar}
\ee
Using (\ref{c-gen},\ref{zeta-gen}) and replacing $\tilde\gamma(\rho)$ by its background value $\gamma=\tilde\gamma(\rho_{0})$ (\ref{gamma0}) we obtain
\bea
	c &=& \frac{\hbar}{m}\pi\rho
	\left[1-\frac{1}{\gamma}
	+...\right] \, ,
 \label{cLD} \\
	\zeta &=& \frac{\hbar}{m}\pi
	\left[1-\frac{3}{2\gamma}
	+...\right] \, ,
 \label{zetaLD}
\eea
where for the sake of brevity we kept only first couple of terms of the corresponding expansions.

A derivation of the dispersive term $A(\rho)$ requires somewhat more involved analysis and will be considered elsewhere. The value $A$ vanishes in the limit $\gamma\to\infty$ (see below).

The limit of an extremely strong coupling $\gamma\to \infty$ is known as the Tonk's limit.
This limit corresponds to impenetrable bosons which in turn can be mapped to free one-dimensional fermions (see Refs.~\onlinecite{FBMT,olshani} for recent discussions).
The collective description in this limit is essentially given by $A(\rho)=0$, $\eta_{B}=0$ and
\begin{eqnarray}
	w(\rho)&=&\frac{\pi^2 \hbar^2\rho^2}{2m^2}\, ,
 \la{wrho1f} \\
 	c &=& \frac{\hbar}{m}\pi \rho_{0} \,.
\end{eqnarray}
The effective KdVB equation becomes the Riemann-Hopf equation (Eq.~\ref{RHeq}) with
\bea
	\zeta=\frac{\hbar}{m}\pi.
\eea
In fact, for 1D fermions the left and right moving waves are decoupled and the equation (Eq.~\ref{RHeq}) follows directly and exactly from the Euler and continuity equations with (\ref{wrho1f}).

An important point about the Tonks limit lies in the absence\cite{BD,BDtonks,gr} of the quantum pressure or the zero point fluctuation term (i.e., $A=0$, $\alpha=0$).
Therefore, in the Tonks limit one does not expect to see the dispersive KdV-like oscillations. To account for a dissipative mechanism that might be present in experiments, the dissipative term term $\nu u_{\xi\xi}$ can be added with the effective equation becoming the Burgers equation (\ref{burgers}).

\subsection{Fermi gas at unitarity}
\label{UFG}

\begin{figure}
\begin{center}
\includegraphics[scale=.57,angle=0]{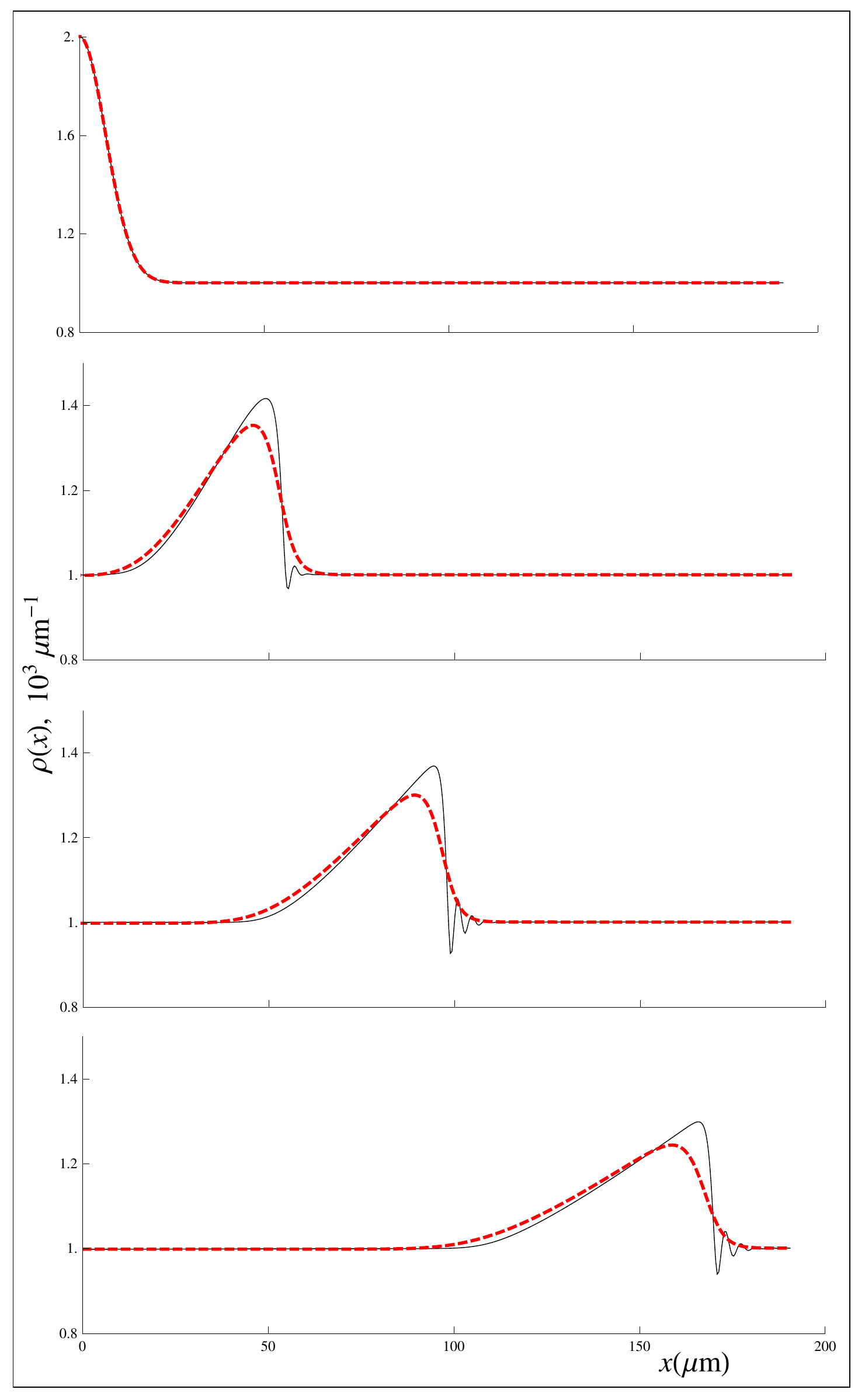}
\caption{
Numerical density evolution (based on smoothed-particle approach\cite{sph,sph1,jjka,jjka1}) of a Gaussian initial profile (with profile scales similar to the experiment in Ref.~\onlinecite{jjka}) in a Unitary Fermi gas in cylindrical geometry (see Sec \ref{UGq1dtrap-estimates} for details). We use the symmetry of the plot and show only its positive x-axis part for visual clarity.  The red (dashed) plot depicts a situation when the dissipative term dominates the dispersive one \cite{jjka,jjka1} ($\alpha_{\eta} = 1$) and the black plot represents a situation when both terms are important ($\alpha_{\eta}=0.1$). Albeit in highly nonlinear regimes the plots shown can be qualitatively described by KdVB physics of Fig.~\ref{estimate}. Indeed, the black plot is similar to the one of Fig. \ref{fig_kdvb_dsw_compact} while the red (dashed) one is reminiscent of the evolution governed by the Burgers equation. See Sec.~\ref{UFG} for related discussions.
}
\label{fig_comp}
\end{center}
\end{figure}

Recently there have been numerous experiments studying the dynamics of strongly interacting cold Fermi atoms and, in particular, of the Fermi gas at unitarity \cite{hara-science,CaoViscosity,KinastDampTemp,JosephSound}. The Fermi gas at unitarity is highly hydrodynamic even deeply in the nonlinear regime\cite{jjka}. Interacting fermions can also be studied in a quasi-one-dimensional regime where the effective 1D hydrodynamics is obtained as a result of the dimensional reduction of the 3D hydrodynamic equations \cite{jjka,luca}.

In this section we consider the superfluid hydrodynamics of the Fermi gas at unitarity. We assume that the temperature is low so that the gas can be effectively described as a simple fluid instead of two-fluid hydrodynamics relevant for superfluids at finite temperatures. We do take the presence of the normal component of the fluid into account only through the effective shear viscosity which is of the order of $\eta_{shear}\sim \frac{\hbar}{m}\rho$ with the coefficient of the order of one under typical experimental conditions\cite{jjka}. The bulk viscosity should be absent in the hydrodynamics of the gas due to the scaling invariance at unitarity\cite{son}.

In the following we consider two types of one-dimensional behavior of nonlinear waves in unitary Fermi gases. The first one is the plane wave propagation through the 3D Fermi gas (considered in Sec.~\ref{UGq1d}) and the other one is the propagation of waves in the Fermi gas confined to elongated cigar-shaped traps where the effective 1D reduction of 3D hydrodynamics is used \cite{jjka} (see Sec.~\ref{UGq1dtrap}). The results are summarized in the Table \ref{table_values}.

\subsubsection{Plane waves in 3D unitary gas}
\label{UGq1d}

The 3D superfluid hydrodynamics of a unitary Fermi gas is given by the continuity and the Euler equations with the chemical potential fixed by the scaling invariance as
\bea
	w(\rho)= (1+\beta)\frac{\hbar^{2}}{2m^2}\left(3\pi^{2}\rho\right)^{\frac{2}{3}}  \, ,
 \label{wpw}
\eea
where $\beta\approx -0.61$ is the Bertsch parameter\cite{OHaraScience,Heiselberg,ThermoLuo}.
Here the density is the 3D density of the Fermi gas. In addition, we take into account the gradient corrections coming from the Madelung's pressure term with\cite{afoot}
\be
	A(\rho)=\frac{\hbar^{2}}{2m^{2}}
 \la{A3Dunitary}
\ee
and the effective shear viscosity $\eta_{shear}$ which appear as a result of the presence of the normal component.

We consider the plane wave solutions of the corresponding hydrodynamic equations assuming that the fluid moves along $z$ direction and
\bea
	\rho(x,y,z) &=& \rho(z),
 \nonumber \\
 	v_{z}(x,y,z) &=&  v(z) \, .
 \la{1dred}
\eea
The substitution of (\ref{1dred}) into 3D hydrodynamic equations
results in 1d equations (\ref{cont},\ref{euler}) with
the effective 1D bulk viscosity $\eta_{B}$ originating from the shear viscosity of the 3D fluid
\bea
	\eta_{B}=\frac{4}{3}\eta_{shear}.
\eea
We obtain the effective parameters of KdVB as
\be
	c =\left(\frac{1+\beta}{3}\right)^{1/2}\frac{\hbar}{m}(3\pi^{2}\rho_{0})^{1/3}
\la{c_pw}
\ee
and
\bea
	\zeta &=&\frac{2}{3}\frac{c}{\rho_{0}} \, ,
 \la{zeta3Du}\\
 	\alpha &=&\frac{\hbar^{2}}{8m^{2}c} \, ,
 \la{alpha3Du}\\
 	\nu&=&\frac{2}{3}\alpha_{\eta}\frac{\hbar}{m} \, .
 \la{nu3Du}
\eea
We notice here that the effective dissipation parameter $\nu$ is defined by the shear viscosity of 3D unitary gas $\eta_{shear}=\alpha_{\eta}\frac{\hbar}{m}\rho_{0}$. Here $\alpha_{\eta}$ is the dimensionless shear viscosity. Putting $\alpha_{\eta}$ to zero would lead to dispersive shock waves (see e.g., Ref.~\onlinecite{luca}). However, the shear viscosity of the unitary gas has a non-trivial temperature behavior.\cite{enss,SchaferViscosity} It has a universal dependence $\alpha_{\eta}\sim T^{3/2}$ at large temperatures and diverges as $\alpha_{\eta}\sim T^{-5}$ at very low temperatures. The latter divergence is related to the divergence of a phonon mean free path (determined by the rate of phonon-phonon collisions) at low temperatures. \cite{lk} As a result the shear viscosity of a unitary gas is expected to develop a minimum at the temperature close to the temperature of a superfluid phase transition with $\alpha_{\eta}\sim 0.5$ \cite{enss,SchaferViscosity}.

Substituting (\ref{zeta3Du},\ref{alpha3Du},\ref{nu3Du}) into the criterion for dispersive shock waves  (\ref{Nosc}) we obtain $\alpha_{\eta} \ll  (3U/(16\rho_{0}))^{1/2}$ which is not possible even extrapolating the amplitude  of the wave into deeply nonlinear regime $U\sim \rho_{0}$. We will see below that the unitary gas in a quasi-1D trap gives a more feasible way to observe dispersive shock waves.

\subsubsection{3d Unitary gas in quasi-1d trap}
\label{UGq1dtrap}

There is another important case when the dynamics of a unitary Fermi gas can be reduced to the one-dimensional. This is the case of a unitary gas confined to elongated cigar-shaped traps. In this case the longitudinal waves along the trap can be described by the effective reduced 1D hydrodynamics \cite{jjka}. It is convenient to use the effective one-dimensional density $\rho(x,t)$ equal to the 3D density integrated over the transverse section of the trap. Then the effective hydrodynamics can be casted in the form of Eq. \ref{cont} and Eq. \ref{euler}, where\cite{jjka,jjka1}
\begin{eqnarray}
	w(\rho)&=& \frac{\hbar^{2}}{2m^{2} l_{\perp}^{2}}(1+\beta)
	\left(\frac{15\pi}{2} \frac{\rho \, l_\perp}{1+\beta} \right)^{2/5} 
 \label{wro_q1d}
\end{eqnarray}
with the transverse oscillator length given by the radial trap frequency $\omega_{\perp}$
\be
	l_\perp=\sqrt{\frac{\hbar}{m\omega_{\perp}}}\, .
 \la{lperp}
\ee
The Madelung term of (\ref{euler}) is given by the reduction of the bulk Madelung term (\ref{A3Dunitary}) and turns out to be\cite{afoot}
\bea
	A(\rho)=\frac{9}{20}\frac{\hbar^2}{m^{2}} \, .
 \la{A1Dunitary}
\eea
The effective one-dimensional bulk viscosity coefficient $\eta_{B}$  originates from the normal component of the superfluid. It should be obtained in the process of averaging of the effective shear viscosity of the 3D superfluid unitary gas. It is also affected by the boundary of the trap where hydrodynamic approximation does not work and a more complete kinetic theory should be considered. This derivation is beyond the scope of this paper and here we treat $\eta_{B}$ as a phenomenological parameter. At the typical conditions of the existing experiments \cite{jjka} it is
\bea
	\eta_{B}(\rho) = \alpha_{\eta} \frac{\hbar}{m}\rho \,,
 \label{unitary_visc}
\eea
where $\alpha_{\eta}$ is a dimensionless parameter of the order of 1. It has been estimated in Ref.~\onlinecite{jjka} as being in the range $1-10$ for the conditions of Ref.~\onlinecite{jjka}.

It was shown in Ref.~\onlinecite{jjka} that the hydrodynamics (\ref{cont},\ref{euler}) with (\ref{wro_q1d},\ref{unitary_visc}) gives a very good quantitative description of the cloud collision experiment. The dispersive term was neglected in Ref. \onlinecite{jjka} as it turns out to be inefficient under the experimental conditions (see below). The hydrodynamic equations can not be solved analytically and numerical solutions were used in Ref. \onlinecite{jjka}. In the following we use instead the mapping to the effective KdVB equation to identify important scales in the problem and to analyze the feasibility of observing the dispersive effects in cold Fermi atoms.

Having the unitary gas dynamics in the form (\ref{cont},\ref{euler}) we immediately derive the parameters of the effective KdVB equation (\ref{kdvb}) in the limit of weak nonlinearity, dispersion and dissipation. The sound velocity is given by (\ref{c-gen}) and is equal to
\bea
	c=\left(\frac{1+\beta}{5}\right)^{1/2}
	\frac{\hbar}{ml_{\perp}} \left(\frac{15\pi}{2} \frac{\rho_{0} l_\perp}{1+\beta} \right)^{1/5} \,.
 \la{c1Dunitary}
\eea
Then (\ref{zeta-gen},\ref{alpha-gen},\ref{nu-gen}) give
\bea
	\zeta&=&\frac{3c}{5\rho_{0}} \, ,
 \la{zeta1Dunitary} \\
 	\alpha&=&\frac{9\hbar^{2}}{80m^{2}c} \, ,
 \la{alpha1Dunitary} \\
 	\nu&=& \alpha_{\eta} \frac{\hbar}{m} \, .
 \la{nu1Dunitary}
\eea

\subsubsection{Shock waves in unitary gas in quasi-1d trap}
\label{UGq1dtrap-estimates}

Let us now use the effective KdVB equation for the unitary Fermi gas in the elongated harmonic trap to understand the qualitative behavior and relevant scales for particular experimental conditions of Ref.~\onlinecite{jjka}. In the latter experiment the two-component Lithium gas was cooled in the trap with the radial frequency $\omega_{\perp}=2\pi \times 437 Hz$. The corresponding transverse oscillator length (\ref{lperp}) can be easily found as
\be
	l_{\perp} = 1.8 \mu m \,,
\ee
using $m_{Li}=1.15 \times 10^{-26}$\, kg corresponding to $\hbar/m_{Li}\approx 9.2$\,$\mu$m$^{2}$/ms. The one-dimensional density in the middle of the trap is
\be
	\rho_{0} \approx 1.1\times 10^{3}\mu m^{-1}\,,
 \la{rho01Dunitary}
\ee
and could be determined from the equation of the state of the unitary Fermi gas, the trap parameters and the total number of particles ($N\sim 2\times 10^{5}$). The temperature in the experiment was very low and will be considered as zero here (except in using the effective viscosity parameter).

The experiment\cite{jjka} studied the collision of two atomic clouds created within the trap using the blue detuned laser beam. Here, instead we make our estimates for a weakly nonlinear, dispersive and dissipative limit presumably described by the effective KdVB equation (\ref{kdvb}) with the speed of sound given by (\ref{c1Dunitary}) and other parameters by (\ref{zeta1Dunitary},\ref{alpha1Dunitary},\ref{nu1Dunitary}) as
\bea
	c &\approx& 14.5\frac{\mu m}{ms}\,,
 \la{c_duke}
 \\
	\zeta &\approx& 8\times10^{-3}\frac{\mu m^{2}}{ms}\,,
 \\
	\alpha &\approx& 0.65\frac{\mu m^{3}}{ms}\,,
 \\
	\nu &\approx& \alpha_{\eta}\times9.1\frac{\mu m^{2}}{ms}
	\quad \mbox{with }\; \alpha_{\eta}\approx 10\,.
\eea

Let us now assume that we created a step-like profile of the density of the typical scale of $U\sim\epsilon \rho_{0}$ or (\ref{rho01Dunitary})
\bea
	U \sim \epsilon\times 1.1\times 10^{3}\mu m^{-1}\,,
\label{Ueps}
\eea
where we keep $\epsilon \ll 1$ as a parameter.
Then the relevant spatial scales (\ref{Wscales}) are estimated as
\bea
	W_{\zeta\nu}&\sim& \alpha_{\eta}\epsilon^{-1}\;1\mu m,
 \\
 	W_{\zeta\alpha}&\sim& \epsilon^{-1/2}\;10^{-1}\mu m,
 \\
 	W_{\nu\alpha}&\sim& \alpha_{\eta}^{-1}\;10^{-1}\mu m.
\eea
We find that the condition (\ref{Nosc}) becomes
\bea
	N = \frac{W_{\nu\alpha}}{W_{\zeta\alpha}} \sim \epsilon^{1/2}\alpha_{\eta}^{-1}
	\gg 1
\eea
and is definitely not valid for $\alpha_{\eta}$ of the order of one or larger.
Therefore, unless the dimensionless viscosity $\alpha_{\eta}$ is somehow experimentally suppressed, the shock waves have the conventional dissipative character and dispersive shock waves should not be observed.

Indeed, the steep density profiles in the experiment of Ref.~\onlinecite{jjka} were identified as the dissipative shock waves. Stretching our estimates to $\epsilon\sim 1$ to take into account the highly nonlinear character of the cloud collision in the experiment (of course, KdVB derived in the limit of weak nonlinearity can serve here only as a tool for estimates) we obtain that $N\sim 0.1$ and the approximate width of the dissipative shock front should be of the order of
\bea
	W_{CSW} \sim W_{\zeta\nu}\sim 10\mu m.
\eea
This number is indeed of the order of the one observed experimentally\cite{jjka}.

To see whether it is feasible at all to observe the dispersive shock waves in cold Fermi atoms we make the following crude estimate. We start with the condition (\ref{Nosc}) rewritten as
\be
	\nu^{2} \ll \alpha \zeta U
\ee
and use for the estimate $\alpha\sim \frac{\hbar^{2}}{m^{2}c}$, $\nu\sim \alpha_{\eta}\frac{\hbar}{m}$, $\zeta\sim \frac{c}{\rho_{0}}$ and $U\sim \rho_{0}$. We have
\be
	\alpha_{\eta} \ll 1
 \la{alphaetaineq}
\ee
as a necessary condition for the observation of the oscillations in the stationary shock wave profile.
Is it possible to lower the effective viscosity to achieve these conditions in the experiment similar to Ref. \onlinecite{jjka}? A straightforward estimate of the mean free path for phonon-phonon collisions for conditions of Ref. \onlinecite{jjka} gives
\be
	\lambda  \sim (T_{F}/T)^{9}\; 4\times 10^{-8}\,\mu\mbox{m} \,.
 \la{lambda}
\ee
We used here the expression\cite{SchaferViscosity} $\lambda= \frac{\hbar c}{E_{F}}\times2.8\times10^{-5}(1+\beta)^{5}\left(\frac{T_{F}}{T}\right)^{9}$ with $T_{F}\approx 1.1\,\mu\mbox{K}$ and $c=v_{F}/\sqrt{3} \approx 19\,\mu\mbox{m/ms}$ calculated at the center of the trap corresponding to a 3D density\cite{jjka} $n_{3D}=6.1 \mu m^{-3}$.

From (\ref{lambda}) we see that the mean free path evaluated at the center of the trap becomes of the order of the trap size ($\sim 200\,\mu$m from center to the edge of the trap) already at $T\approx 0.1T_{F}$ which is roughly the temperature of the cold atom system in Ref. \onlinecite{jjka}. Therefore, we expect that if one lowers the temperature the mean free path will saturate at the size of the trap and the shear viscosity of 3D superfluid will be dominated by the density of the normal component of the superfluid $\rho_{n}\sim T^{4}$ (from kinetic theory $\eta_{shear}=\rho_{n}p\lambda$,  where $p$ is the average momentum).
Now it is clear that the decrease of the temperature by the factor of 2 in experiment \cite{jjka} might decrease the normal density and therefore, shear viscosity at the center of the trap by a factor of 20. Although the effective 1D viscosity should be obtained as a result of a complicated averaging of the two-fluid hydrodynamic equations in directions transverse to the trap this rough estimate shows that the possibility to observe the dispersive behavior in quasi 1D traps is indeed feasible and requires the experiments at somewhat lower temperature than the one in Ref.~\onlinecite{jjka}.

To show that the presented analysis of the shock wave physics of KdVB equation describes qualitatively the solution of actual hydrodynamic system (\ref{cont},\ref{euler}) we present the numerical solution of hydrodynamic equations in Fig.~\ref{fig_comp} for the hydrodynamic system derived for a unitary gas in a quasi-1d trap (see Ref. \onlinecite{jjka} and the section \ref{UGq1dtrap} of this paper).

We obtain numerical solutions of the continuity and Euler equations
using the technique of smoothed-particle hydrodynamics\cite{jjka,jjka1,sph,sph1}. This numerical method is a mesh-less algorithm. The algorithm essentially consists of mapping the system of differential equations describing density and velocity fields to dynamic equations of a sufficiently large set of moving pseudo-particles (approximating the density and velocity fields) and then solving the molecular dynamics of these pseudo-particles.

The results of a numeric solution of hydrodynamic equations are shown as a red (dashed) plot in Fig.~\ref{fig_comp}. The parameters in equations are taken similar to the conditions of the experiment \cite{jjka} discussed in the section \ref{UGq1dtrap-estimates}. We used a constant background with the background density given by Eq.~\ref{rho01Dunitary} with the initial density profile of a Gaussian form of the height equal to $\rho_{0}$, ie, in Eq.~\ref{Ueps} we take $\epsilon = 1$. The full-width at half maxima of the Gaussian bump is taken to be about $20\,\mu$m (5\% of system size) similar to the experiment \cite{jjka}. The red (dashed) plot in Fig.~\ref{fig_comp} corresponds to $\alpha_{\eta} = 1$ and we clearly see that the shock waves are of dissipative nature. Qualitatively the red (dashed) plot is very similar to the evolution described by the Burgers equation (\ref{burgers}). We also plot (black plot) the profile obtained numerically for $\alpha_{\eta}=0.1$ (see the discussion of the feasibility of obtaining small viscosities in finite systems above). For this value of effective dissipation we clearly see non-vanishing oscillations in the stationary profile of the shock wave.

\section{Conclusion}
\label{conclusion}

We presented a unified picture of dynamics of a large class of nonlinear systems in the limit of weak nonlinearity, dissipation and dispersion (WNDD). This picture is delineated by the Korteweg--de Vries--Burgers equation (\ref{kdvb}) for the deviation of the density from its background value. We described various regimes of nonlinear evolution described by KdVB and summarized these regimes in the diagram in Figure \ref{estimate}. The parameters of KdVB have been related to those of a generic one-dimensional fluid and then to particular parameters of several models used in cold atom research. These relations are summarized in the Table \ref{table_values}.

Arguably the most interesting regime in WNDD limit is the formation of dispersive shock waves which are known to occur in interacting systems\cite{BD,AblowitzEtAl,key-8wiegmann}. The analysis presented in this paper uses the well-known results in the theory of nonlinear partial differential equations. It allows to make quick estimates of important scales and parameters of shock waves.

Although in this work we focused on a Galilean invariant fluid, the general analysis can be extended to other systems as well. KdVB equation still remains the most universal description in WNDD limit unless additional symmetries restrict the form of the terms of the lowest order in nonlinearity and/or field gradients (e.g., the underlying particle-hole symmetry might require the symmetry $u\to-u$ so that the lowest allowed nonlinear term has a form of $u^{2}u_{x}$ leading to the so-called modified KdV equation). The WNDD limit therefore allows to divide the complicated dynamics of nonlinear systems into several distinctive ``nonlinear universality classes'' giving an insight into their generic dynamic features (for a recent related discussion see Refs.~\onlinecite{eugene2012,aditi_xxz,aditi_domain_wall}).

We have also discussed another important way of generalizing KdVB description appropriate for systems with long-range interactions. For those systems the locality of the effective description is not required. The relaxation of the locality requirement leads to a wider class of integro-differential effective equations (see Appendix \ref{long-range-app} for some examples).

Another important assumption used in this paper is the possibility to have a consistent effective description of the system by a single-component fluid. Of course, the presence of several components such as normal/superfluid components in BECs or equivalent components in spinor BECs can be crucial for the accurate description of corresponding physical systems.

The described reduction of nonlinear three-dimensional dynamics to effective chiral one-dimensional equation obtained in WNDD limit (\ref{kdvb}) cannot capture all fascinating and complex effects in collective behavior of interacting 3d systems. For example, taking into account the next order terms in the reductive perturbation expansion will lead to the interaction between chiral sectors (left and right moving waves) neglected in this work. The next order terms in time derivatives will result in the frequency dependence of kinetic coefficients essential for e.g., the physics of the unitary gas\cite{enss,edt}.

Finally, the description used in this work is completely classical (for example, there is no $\hbar$ in Secs.~II-V. The quantum physics enters here only through the values of effective parameters of classical hydrodynamic equations (see Sec.~\ref{examples}). One of the most interesting problem is to identify the quantum effects not describable by conventional classical physics (quantization of vorticity in superfluid system can serve as an example of such effects).

\section{Acknowledgments}

We are grateful to J. E. Thomas,  J. Joseph, D. H. J. O ' Dell, J. Thywissen, M. Punk and E. Taylor  for useful discussions. We thank the Princeton Center for Theoretical Science for their hospitality during the workshop on \textit{Ultracold Atoms and Magnetism} where this work started. The work of A. G. A. was supported by the NSF under Grant No. DMR-0906866. M. K. thanks the hospitality of the Canadian Institute for Advanced Research (CIFAR) funded \textit{Cold Atoms Meeting} at Banff where several interesting discussions took place.

\appendix

\section{Relaxing locality}
\label{long-range-app}

In the main part of the paper we assumed that the system of particles is well described by a local hydrodynamics theory. The assumption of locality was important in deriving the effective KdVB equation for those systems. In the presence of long range interactions between particles our derivations should be modified and generally different counting scheme should be used in the reductive perturbation method. In this appendix we show how the assumption of locality can be relaxed on the example of two models: the Calogero model and the dipolar BEC model. The former is often used as a theoretical toy model as it is exactly integrable and very well studied. The latter is relevant to some experimental cold atom systems\cite{axelexp,dy}.

\subsection{Calogero Model}
\label{CM}

The Calogero model is the model of one-dimensional particles interacting via the inverse square potential\cite{Sutherland-book}. The Calogero Hamiltonian is given by
\begin{eqnarray}
    H &=& \frac{1}{2m}\sum_{j=1}^{N}p_j^{2}
    + \frac{\hbar^2}{2m}\sum_{j,k=1; j\neq k}^{N} \frac{\lambda(\lambda-1)}{(x_{j}-x_{k})^{2}}.
\label{cal_ham}
\end{eqnarray}
Here the coupling constant $\lambda$ is dimensionless. For simplicity we consider here the quasiclassical limit corresponding to $\lambda\gg 1$ and in the following replace $\lambda(\lambda-1)\to \lambda^{2}$ and neglect corrections of the order of $1/\lambda$.  It turns out that one can write down the collective field theory of the above model exactly\cite{JevickiSakita,Jevicki-1992,2009-AbanovBettelheimWiegmann,2005-AbanovWiegmann,1995-Polychronakos}, ie, including all gradient and nonlinear terms. This field theory can be again casted in the form (\ref{cont},\ref{euler}) with no dissipation $\eta_{B}=0$ and
\begin{eqnarray}
	w(\rho)&=&\frac{\hbar^{2}\lambda^{2}}{m^{2}}
	\left(\frac{1}{2}(\pi  \rho)^{2}+\pi \rho_{x}^{H}\right)\, ,
 \la{w-calogero}\\
 	A(\rho)&=&\frac{\hbar^2\lambda^{2}}{2m^2}\,.
\end{eqnarray}
Here the superscript $H$ stands for the Hilbert transform defined by
\begin{equation}
 \label{hilbert}
	f^{H}(x)=\frac{1}{\pi}P.V. \int\frac{f(y)\, dy}{y-x}.
\end{equation}
Notice that the second term in (\ref{w-calogero}) is nonlocal reflecting the long range interaction between Calogero particles in Eq.~\ref{cal_ham}.

The sound velocity in the collective field theory for Calogero model \cite{JevickiSakita,Jevicki-1992,2009-AbanovBettelheimWiegmann,2005-AbanovWiegmann} can be obtained from (\ref{w-calogero}) using (\ref{c-gen}) and is given by
\bea
	c = \frac{\hbar}{m} \lambda \pi\rho_0\, .
\eea
The second term in (\ref{w-calogero}) is of the first order in $\partial_{x}$ and the Burgers counting scheme (\ref{bur_rho}-\ref{bur_per_g}) should be used in the reductive perturbation method of Sec.~\ref{sec:redpert}. As a result we obtain the so-called Benjamin-Ono equation
\begin{equation}
 \label{BO_eq_example}
	u_t+\zeta uu_{\xi}+ \alpha_{BO}\left(u_{\xi\xi}\right)^{H}
	=0\qquad\mbox{\textbf{(Benjamin-Ono)}}
\end{equation}
as an effective description of Calogero hydrodynamics in the limit of weak nonlinearity and dispersion\cite{2009-AbanovBettelheimWiegmann,2005-AbanovWiegmann}.
The coefficients of (\ref{BO_eq_example}) are given by
\bea
	\zeta&=&\frac{\pi\hbar\lambda}{m} \, ,
 \\
 	\alpha_{BO}&=&\frac{\hbar \lambda}{2 m} \, .
\eea
Two remarks are in order. First, notice that the Benjamin-Ono term (the last term of Eq.~\ref{BO_eq_example}) dominates the conventional KdV term $u_{\xi\xi\xi}$ in the long wave limit. It is of the same order as the Burgers term $u_{\xi\xi}$. Second, although the term $u_{\xi\xi}^{H}$ scales similar to the Burgers term, it is very different in nature as it does not result in any dissipation. Indeed, the Fourier transform $(u_{\xi\xi}^{H})_{k}=ik|k|u_{k}$ and this term results in the real  correction $\omega \sim \alpha_{BO} k|k|$ to the spectrum of linear waves in contrast to the Burgers term giving imaginary (dissipative) correction $\omega\sim -i\nu k^{2}$.

\subsection{Quasi-1D Dipolar BEC}
\label{dipolar-bec}

Systems of bosons interacting via long range potential have been recently realized experimentally\cite{axelexp,dy} for bosons with dipole interactions. The quasi-1D dipolar BEC can be described by a non-local version\cite{duncan,duncan1,axel} of the Gross-Pitaevskii\cite{becrmp} equation. By choosing a sufficiently large radial trap frequency it is possible to freeze the radial motion\cite{cai2010,duncanEPJ} of the BEC and thereby obtain a quasi-1D dipolar BEC described by the following non-local one-dimensional GPE (see Ref. \onlinecite{cai2010} for the derivation)
\begin{eqnarray}
\label{dipolar_gpe1}
	i \hbar\partial_{t}\psi(x,t)
	&=&\left(-\frac{\hbar^2}{2m}\partial_{x}^{2}
	+m w(|\psi|^{2})\right)\psi(x,t)\,,
\end{eqnarray}
where
\be
	w(\rho) = \frac{g}{m l_{\perp}^{2}}\left(A_{1D}\rho
	+B l_{\perp}^{2}\partial_{x}^{2}\,\rho^{U}\right)\,.
 \la{wdipolar}
\ee
Here  $g$ is the 3D contact interaction strength,
$l_{\perp,z}=\sqrt{\hbar/(m\omega_{\perp,z})}$ is the transverse (axial) oscillator length, and the dimensionless constants $A_{1D}$ and $B$ are
\bea
	A_{1D} &=& \frac{1}{2\pi}
	\left[1+\frac{1}{2}\epsilon_{dd}\left(1-3n_{z}^{2}\right)\right] \, ,
 \\
 	B &=& \frac{3}{8\pi}  \epsilon_{dd}\left(1-3n_{z}^{2}\right) \, .
\eea
$n_z$ is the $z$ component of the direction of the dipole axis $\hat{\mathbf{n}}$ and $\epsilon_{dd}$ is a natural dimensionless parameter for the relative strength of dipolar and s-wave interaction\cite{cai2010}.

In (\ref{wdipolar}) we used the dipolar integral transform
\bea
	f^{U}(x) = \int_{-\infty}^{+\infty}
	U_{1D}(x-y)f(y)\,dy
 \la{Utransform}
\eea
defined by its  kernel
\bea
 \label{erfc}
	U_{1D}(x)=\sqrt{\frac{\pi}{2\,l_{\perp}^{2}}}\,
	e^{\frac{x^{2}}{2 l_{\perp}^{2}}}\mbox{erfc}\left(\frac{|x|}{\sqrt{2} l_\perp}\right),
\eea
where erfc is the complementary error function.
The Fourier transform of $U_{1D}(x)$ is given by
\bea
	\tilde{U}_{1D}(k)
	=\int_{0}^{\infty}d\kappa \,\frac{e^{-\kappa/2}}{\kappa+(kl_{\perp})^{2}}
\eea
and its asymptotic behavior as $k\rightarrow 0$ is
\bea
	\tilde{U}_{1D}(k)\sim
	-\gamma_{E}-\ln\left(\frac{(kl_{\perp})^{2}}{2}\right)
	+\ldots\,,
 \la{Ukto0}
\eea
where $\gamma_{E}\approx 0.5772$ is  the Euler-Mascheroni constant.

The hydrodynamic change of variables $\psi=\sqrt{\rho}e^{i\frac{m}{\hbar}\int_{0}^{x}v(x^\prime)dx^\prime}$ brings (\ref{dipolar_gpe1}) to the hydrodynamic form\cite{duncanEPJ} (\ref{cont},\ref{euler}) with the chemical potential $w(\rho)$ given by (\ref{wdipolar}) and
\bea
	A(\rho)&=&\frac{\hbar^{2}}{2m^{2}}.
\eea
The speed of sound in this system can be easily found using (\ref{c-gen},\ref{wdipolar})
\be
	c = \sqrt{\frac{g\rho_{0}}{ml_{\perp}^{2}}A_{1D}}
	=\frac{\hbar}{ml_{\perp}}(\gamma A_{1D})^{1/2},
\ee
where we introduced the dimensionless coupling constant
\be
	\gamma = \frac{m\rho_{0}}{\hbar^{2}}g.
\ee
The resulting Euler equation in this case is non-local.
The perturbative scaling scheme (Eqns. \ref{kdvb_rho},\ref{kdvb_v},\ref{pertf},\ref{pertg}) gives rise to the following non-local-KdV-like equation (``dipolar-KdV'')
\begin{equation}
 \label{nkdv_eq_example}
	u_t+\zeta uu_{\xi}- \alpha u_{\xi\xi\xi} +\alpha_{D}u_{\xi\xi\xi}^{U}=0
	\qquad\mbox{\textbf{(dipolar-KdV)}}
\end{equation}
with
\bea
	\zeta&=&\frac{3}{4}\frac{c}{\rho_{0}} \,,
 \\
	\alpha &=&\frac{\hbar^2}{8m^2 c}\, ,
 \\
	\alpha_{D}&=& \gamma B  \frac{\hbar^{2}}{2m^{2}c} \,.
\eea
Here the superscript $U$ again denotes the transform (\ref{Utransform}). As we work only in the long wavelength limit we can think of the transform's kernel as of Fourier transform of (\ref{Ukto0}). Although, technically in the asymptotic long  wavelength limit $k\to 0$ the $\alpha_{D}$ term will dominate the $\alpha$ term we keep both of them in (\ref{nkdv_eq_example}). Indeed, the kernel (\ref{Ukto0}) is growing logarithmically slow with $k$ and both terms become of the same order at $kl_{\perp}\sim \exp(\pi\alpha/\alpha_{D})$ which is defined by the actual values of the dimensionless parameters of the dipolar system.


\bibliographystyle{my-refs}
\bibliography{wndd}

\begin{thebibliography}{10}
\bibitem{afoot}
We remark here that one can trust the value (\ref{A3Dunitary}) and later the
  value of (\ref{A1Dunitary}) only by the order of magnitude. The Madelung's
  pressure term is higher order in hydrodynamic gradient expansion and can, in
  principle, have contributions additional to the simplest form
  (\ref{A3Dunitary}) obtained from the Gross-Pitaevskii equation.
\end{thebibliography}

\end{document}